\documentclass[superscriptaddress,amssymb,amsmath]{revtex4}
\usepackage{natbib,graphicx,subfigure,subeqnarray,fancyhdr,amsmath,epstopdf,bm}

\begin{document} 
\newcommand{\Fc}{\mathcal{F}}
\newcommand{\Rc}{\mathcal{R}}
\newcommand{\dd}{\mathrm{d}}
\newcommand{\ee}{\mathrm{e}}
\newcommand{\ci}{\mathrm{i}}
\newcommand{\ib}{\mathbf{i}}
\newcommand{\jb}{\mathbf{j}}
\newcommand{\kb}{\mathbf{k}}
\newcommand{\ab}{\mathbf{a}}
\newcommand{\Fb}{\mathbf{F}}
\newcommand{\fb}{\mathbf{f}}
\newcommand{\Gb}{\mathbf{G}}
\newcommand{\Mb}{\mathbf{M}Ä}
\newcommand{\nb}{\mathbf{n}}
\newcommand{\Sb}{\mathbf{S}}
\newcommand{\ub}{\mathbf{u}}
\newcommand{\Ub}{\mathbf{U}}
\newcommand{\xb}{\mathbf{x}}
\newcommand{\Xb}{\mathbf{X}}
\newcommand{\Rey}{\textit{Re}}
\newcommand{\ddp}{[p]^\pm}
\let\grad\nabla
\newcommand{\z}{\zeta}
\newcommand{\kk}{\kappa}
\newcommand{\tkk}{\tilde{\kappa}}
\newcommand{\dpp}{[p]^\pm}
\newcommand{\e}{\varepsilon}
\newcommand{\zb}{\bar{\zeta}}
\let\grad\nabla
\let\bcdot\cdot
\newcommand{\half}{{\textstyle\frac{1}{2}}}
\newcommand{\textfrac}[2]{{\textstyle\frac{#1}{#2}}}
\newcommand{\LF}[1]{{#1}^{\mathrm{LF}}}
\newcommand{\Lap}[1]{{#1}^{\mathrm{L}}}
\newcommand{\ds}{*\!*}
\newcommand{\cond}[2]{\frac{\mathrm{D} #1}{\mathrm{D} #2}}
\newcommand{\pard}[2]{\frac{\partial #1}{\partial #2}}
\newcommand{\totd}[2]{\frac{\mathrm{d}#1}{\mathrm{d}#2}}
\newcommand{\Real}{\mbox{Re}}
\newcommand{\Imag}{\mbox{Im}}
\newcommand{\Fpint}{=\!\!\!\!\!\!\!\int}
\makeatletter
\def\sgn{\mathop{\operator@font sgn}}
\makeatother
\title{Resonance and propulsion performance of a heaving flexible wing}
\author{S\'ebastien Michelin}
\email{smichelin@ucsd.edu}
\affiliation{Department of Mechanical and Aerospace Engineering, Jacobs School of Engineering, University of California San Diego, 9500 Gilman Drive, La Jolla CA 92093-0411.}
\affiliation{Ecole Nationale Sup\'erieure des Mines de Paris, 60--62 Boulevard Saint Michel, 75272 Paris Cedex 06, France.}
\author{Stefan G. Llewellyn Smith}
\affiliation{Department of Mechanical and Aerospace Engineering, Jacobs School of Engineering, University of California San Diego, 9500 Gilman Drive, La Jolla CA 92093-0411.}
\date{\today}

\begin{abstract}
The influence of the bending rigidity of a flexible heaving wing on its propulsive performance in a two-dimensional imposed parallel flow is investigated in the inviscid limit. Potential flow theory is used to describe the flow over the flapping wing. The vortical wake of the wing is accounted for by the shedding of point vortices with unsteady intensity from the wing's trailing edge. The trailing-edge flapping amplitude is shown to be maximal for a discrete set of values of the rigidity, at which a resonance occurs between the forcing frequency and a natural frequency of the system. A quantitative comparison of the position of these resonances with linear stability analysis results is presented. Such resonances induce maximum values of the mean developed thrust and power input. The flapping efficiency is also shown to be greatly enhanced by flexibility. 

\end{abstract}
\maketitle

\section{Introduction}
Unlike terrestrial animals that can use solid friction and a fixed support, insects and fishes must generate from the surrounding fluid the lift and thrust forces necessary to their motion in their environment  \cite{lighthill1960,lighthill1969,childress1981,triantafyllou2000,wang2005}. Beyond the fundamental interest of understanding the mechanism of insect flight and fish swimming, recent research on propulsion in fluids has also been motivated by the development of micro-aviation vehicles (MAV) and of more efficient propulsion techniques based on biomimetics.

Insects use thin flapping wings to generate an unsteady flow around them to produce these  forces. The flow  is characterized by a relatively large Reynolds number $\Rey=UL/\nu$ with $U$ the typical wing velocity, $L$ its characteristic chord and $\nu$ the kinematic viscosity of the surrounding fluid: Typically, $\Rey\sim 100$--$5000$ \cite{wang2005}; in that range, the forces on the wings are dominated by the pressure contribution and viscous effects are concentrated in thin boundary layers near the solid's boundary. These boundary layers separate during the unsteady wing motion and roll up into strong coherent vortices \cite{thomas2004} that carry momentum away from the insect, thereby generating the propulsive forces. The flow around the insect is highly unsteady, and an ongoing research challenge resides in the ability to describe the forces on the flapping structure without solving explicitely for the details of the flow \cite{ellington1984,dickinson1999,wang2004b,pesavento2004,berman2007}.

An important physical insight into the generation of propulsive forces by flapping or deforming solids has been provided by the study of active motion and deformation. In such experimental \cite{anderson1998,godoydiana2008}, theoretical \cite{wu1961,lighthill1960} and numerical studies \cite{wang2000a,wang2000b,miao2006,shukla2007}, the position of the solid is prescribed and the influence of the swimming stroke on the propulsive performance is analyzed.

The recent development of experimental imaging techniques has however shown that most insect wings are not purely rigid and that their deformation is not entirely controlled by the animal: the wing can experience large passive deformations during the stroke period \cite{wootton1992} under the action of the outside flow and its internal bending rigidy, whose spanwise and chordwise distribution results from the venation pattern of the wing \cite{combes2003a,combes2003b}. The forcing is generally applied by the insect on its wing through a main axis whose rigidity is significantly higher than the rest of the structure (e.g. the leading edge of the wing). An important challenge now is to understand the impact of such passive deformation on the propulsive performance of a flapping structure. In particular, one may be interested in potential reduction of energy usage induced by the flexibility, and examine if the values of the rigidity for these structures lie within an ``optimal" range for which the flapping efficiency is the highest. The definition of optimality in this work in terms only of thrust production and propulsive efficiency is purposely restrictive: from a biological point of view many other factors must be taken into account to define the optimal structure of an insect wing, including but not limited to material resistance and manoeuverability.

The purpose of the present study is to investigate numerically the effect of flexibility on the propulsive characteristics of a flapping appendage. In the following, this structure will be referred to as a wing, understanding that the model could also be applied to a fish fin if it is allowed to deform passively under the effect of the flow and of its bending rigidity. Solving for the coupled motion of a flexible solid and a fluid is computationally challenging and expensive, primarily due to the coupling occurring on a moving boundary whose position is a priori unknown and must be solved for as well. Popular techniques to overcome this difficulty are the use of coupled fluid and solid solvers using fitted grids \cite{connell2007} and immersed boundary methods \cite{zhu2002,zhu2003}. The use of low-order models for the flexible wing also simplifies the computation while still retaining important physical results on the reaction of the body to the fluid flow \cite{bergou2007,toomey2008}.

The model used here focuses on a simplified two-dimensional propulsion problem using a flexible wing of infinite span, actuated at its leading edge in a purely heaving motion and reacting passively to the flow forces and its internal elasticity. The present model does not aim to represent a particular flying or swimming pattern, but rather considers a one-degree-of-freedom forcing to focus on the influence of flexibility on the performance of the apparatus. In the limit of high $\Rey$, viscous forces are neglected and the viscosity's influence on the flow is retained in this potential flow formulation by the irreversible shedding of vorticity from the trailing edge of the flapping structure, in the form of point vortices whose unsteady intensity is determined so as to satisfy the regularity condition at the solid's trailing edge \cite{michelin2009a,michelin2008}.

A similar approach was recently proposed by Alben \cite{alben2008c} for a pitching elastic sheet using a vortex sheet representation of the wake. Optimal values of the solid's rigidity were discussed in the limit of negligible solid inertia (that is particularly relevant in the case of fish swimming) and of infinitesimally small displacements of the solid. The present work builds on these results and considers the general case of non-linear deformations of the sheet with non-negligible inertia (as is the case for a flapping insect wing).  The use of the unsteady point vortex model rather than the full vortex sheet description also allows for a simpler treatment. Similar resonance patterns are observed and a theoretical argument is provided for their origin and position. The relation between thrust or drag production and vortex wake structure is also investigated.

In section \ref{sec:model}, the fluid-solid model is presented and the propulsive performance quantities of interest are defined. Section \ref{sec:numerical} discusses briefly the numerical methods used as well as the existence of a periodic regime. Section \ref{sec:optimal} then investigates the influence of the solid's rigidity on the propulsion forces and efficiency and relates them to the structure of the solid's wake. Peaks of thrust are observed for particular values of the rigidity and in section \ref{sec:resonance}, the occurrence of such peaks is showed to correspond to a resonance between the forcing frequency and the natural frequencies of the fluid-solid system. Finally, section \ref{sec:conclusions} presents some general conclusions and discusses the limitations of the model.

\section{Description of the model}
\label{sec:model}
\subsection{Solid model}
\begin{figure}
\begin{center}
\includegraphics[width=8.6cm]{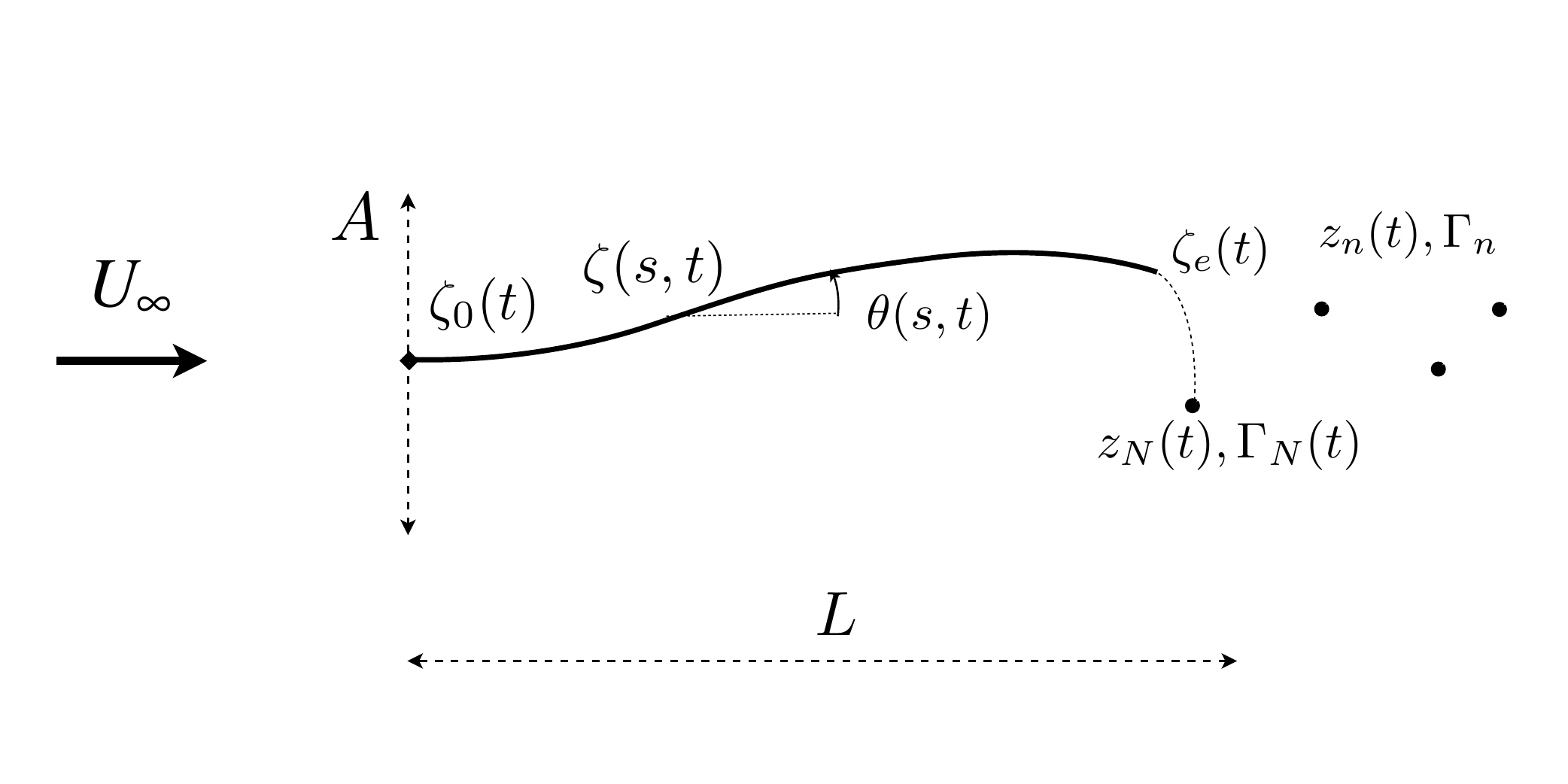}
\caption{Heaving flexible wing in a steady axial flow. The heaving motion of amplitude $A$ is imposed at the leading edge and vortices are shed from the trailing edge.}\label{fig:notations}
\end{center}
\end{figure}

The following two-dimensional model for the flapping structure is considered (see Fig.~\ref{fig:notations}). The wing is represented by an elastic sheet of chord $L$ and infinite span, clamped at its leading edge on an attachment pole of negligible thickness, actuated by the operator (e.g. main body of the insect). The operator applies a purely vertical motion to the sheet's leading edge, whose orientation is constrained to be stricly horizontal. The vertical position and orientation of the wing at the leading edge are then:
\begin{equation}\label{eq:heaving}
h(t)=h_0(1-\cos\omega t)\quad\textrm{and}\quad\theta_0(t)=0,
\end{equation}
so that $A=2h_0$ and $f=\omega/2\pi$ are respectively the amplitude and frequency of the flapping motion (to avoid confusion, $\omega$ or its non-dimensional form will be referred to as the angular frequency in the following). The flapping wing has a chordwise flexural rigidity per unit length $B$ and a mass per unit area $\rho_s$. Its thickness is negligible compared to $L$. The wing is placed in a uniform horizontal flow $U_\infty$ of density $\rho$. The motion of the wing's leading edge is entirely prescribed by \eqref{eq:heaving} but the rest of the wing has a purely passive motion in response to its internal elasticity, the leading-edge forcing and the pressure forces applied by the surrounding flow.

In the following, $L$, $U_\infty$ and $\rho$ are used as reference quantities to non-dimensionalize the problem. The properties of the elastic sheet are characterized by the mass ratio $\mu$ and non-dimensional rigidity $\eta$ defined as
\begin{equation}
\mu=\frac{\rho_s}{\rho L},\qquad \eta=\frac{B}{\rho U_\infty^2L^3},
\end{equation}
and the leading-edge forcing is characterized by the non-dimensional forcing amplitude $\varepsilon$ and frequency $\bar{f}$ 
\begin{equation}
\varepsilon=\frac{h_0}{L}=\frac{A}{2L},\qquad \bar{f}=\frac{fL}{U_\infty}\cdot
\end{equation}
The forcing Strouhal number is defined in accordance with previous experimental studies \cite{anderson1998} as
\begin{equation}
\mbox{\textit{St}}=\frac{fA}{U_\infty}=2\varepsilon\bar{f}.
\end{equation}
The motion of the wing is described using an inextensible Euler--Bernoulli beam representation \cite{michelin2008}. We are interested in large displacements of the wing so all non-linear geometric terms must be included. The linear Euler--Bernoulli assumption remains valid if the curvature radius of the wing is much larger than its thickness, which is assumed here. The position of the wing is described using complex notation as $\z(s,t)=x(s,t)+\ci y(s,t)$ and its orientation is defined as $\theta(s,t)$ with $0\leq s\leq 1$ the curvilinear coordinate along the wing. The classical notation for the complex flow velocity is also used here: $w=u-\ci v$, with $(u,v)$ the cartesian components of the velocity vector. The conservation of momentum for each element of the wing and the inextensibility condition can be written as
\begin{equation}\label{eq:beam}
\mu\ddot{\z}=\left[\left(T-\ci\eta\theta_{ss}\right)\ee^{\ci\theta}\right]_s-\ci[p]^\pm\ee^{\ci\theta},\qquad \z_{s}=\ee^{\ci\theta},
\end{equation}
where $\ddp$ is the pressure difference between the top and bottom sides of the wing and $T$ is the wing tension that must be solved for at each time-step to enforce the inextensibility condition. The clamped-free boundary conditions imposed by the forcing \eqref{eq:heaving} are 
\begin{align}
\z(0,t)&=\z_0(t)=\ci\varepsilon\left[1-\cos(2\pi \bar{f}t)\right],\,\,\, \theta(0,t)=0,\label{eq:bc01}\\
\theta_s(1,t)&=\theta_{ss}(1,t)=T(1,t)=0.\label{eq:bc02}
\end{align}
Equations \eqref{eq:beam}-\eqref{eq:bc02} can be rewritten as a system for $\theta$ and $T$ only \cite{michelin2008}:
\begin{align}
T_{ss}-\theta^2_{s}T&=-\dpp\theta_{s}-2\eta\theta_{s}\theta_{sss}-\eta\theta^2_{ss}-\mu\dot\theta^2\label{eq:solid1}\\
\mu\ddot\theta&=-\dpp_s-\eta\theta_{ssss}+(T+\eta\theta^2_{s})\theta_{ss}+2T_{s}\theta_{s}\label{eq:solid2}\\
\theta(0,t)&=\theta_{s}(1,t)=\theta_{ss}(1,t)=T(1,t)=0\label{eq:bc1}\\
\mu\ddot\z_0+\mu\int_0^1\int_0^s\ee^{\ci\theta}\left(\ci\ddot\theta-\dot\theta^2\right)\dd s'\dd s&=-T(0)+\ci\eta\theta_{ss}(0)-\ci\int_0^1\dpp\ee^{\ci\theta}\dd s.\label{eq:bc2}
\end{align}

\subsection{Representation of the flow around the flapping wing}
The flow around the wing is taken potential. The expected boundary layer separation at the trailing edge and its subsequent roll-up into vortices is represented in this inviscid formulation by a discrete shedding of point vortices from the trailing edge $\z_e=\z(1,t)$ of the flapping wing (Fig.~\ref{fig:notations}). The intensity of the last shed vortex is determined so as to cancel exactly the square-root singularity arising in the velocity field due to the presence of the flat sharp corner \cite{cortelezzi1993,cortelezzi1995,michelin2009a,michelin2008}. When the intensity of the unsteady vortex reaches a maximum, a new vortex is created from the generating corner, thereby expressing the irreversible nature of vortex roll-up in the formalism of this inviscid model. From that point on, the intensity of the previous vortex is frozen. The unsteady point vortices satisfy the modified equation of motion
\begin{equation}\label{eq:BM}
\dot{z}_n+(z_n-\zeta_e)\frac{\dot\Gamma_n}{\Gamma_n}=\overline{\tilde{w}_n},
\end{equation}
where $z_n$ and $\Gamma_n$ respectively refer to the position and intensity of the point vortex. $\tilde{w}_n$ is the desingularized complex velocity at the vortex position and the overbar denotes a complex conjugate. Equation~\eqref{eq:BM} is known as the Brown--Michael equation \cite{brownmichael1954}, and enforces the conservation of fluid momentum around the vortex and associated branch cut in an integral sense \cite{michelin2009a}. The omission of the corrective term on the left-hand-side would lead to an unphysical unbalanced force on the branch cut linking the vortex to its generating corner $\z_e$ \cite{rott1956}.

The shedding of vorticity from the leading edge is neglected here, as we focus mostly on situations where the angle of attack remains small at the leading edge \cite{alben2008c}. Alternatively, this representation can also be seen as the limit case of a smoothed leading edge of very small curvature radius (as in an airfoil profile for example). Only one unsteady vortex is shed at a time (from the trailing edge). Noting $N$ the number of vortices at a particular time, all vortex intensities $\Gamma_n$ are therefore independent of time except for the last one $\Gamma_N(t)$.

In the absence of viscosity, the tangential velocity of the flow can be discontinuous across the wing. The potential flow around the wing is then computed by representing the infinitely thin solid as a bound-vorticity distribution $\kappa$ \cite{jones2003,shukla2007,alben2008}. The complex flow velocity is obtained by superposition of the flow at infinity and the contribution of the bound and wake vorticity
\begin{equation}
\label{eq:vel}
w(z,t)=1+\frac{1}{2\pi\ci}\left[\int_0^1\frac{\kappa\dd s}{z-\zeta(s)}+\sum_{n=1}^{N}\frac{\Gamma_n}{z-z_n}\right].
\end{equation}

The regularity condition at the trailing edge imposes $w(\z_e,t)\neq\infty$ or equivalently $\kappa(1,t)=0$. The total circulation at infinity is conserved and, assuming the system is started from rest, must be zero at all time. The bound-vorticity $\kappa$ is the solution of a singular Fredholm equation obtained by applying the continuity of normal velocity on the wing \cite{shukla2007}. The corresponding system of equations for $\kappa$ and $\Gamma_N$ is
\begin{align}
\frac{1}{2\pi}\int_0^1\Real\left[\frac{\ee^{\ci\theta(s)}}{\zeta(s_0)-\zeta(s)}\right]\kappa(s)\dd s&=\Imag\Bigg[\ee^{\ci\theta}\Bigg(1+\frac{1}{2\pi\ci}\sum_{n=1}^{N}\frac{\Gamma_n}{\zeta-z_n}-\dot{\bar{\zeta}}\Bigg)\Bigg],\label{eq:inteq}\\
\int_0^1\kappa(s)\dd s+\sum_{n=1}^{N}\Gamma_n&=0,\label{eq:kelv}\\
\kappa(1,t)&=0.\label{eq:regul}
\end{align}
From $\kappa$, the pressure jump $\ddp$ across the wing can be computed by integration of Bernoulli's theorem along the wing:
\begin{equation}\label{eq:bernoulli}
\ddp(s_0)=\int_0^{s_0}\dot\kappa(s)\dd s+\kappa(s_0)w_p(s_0)
\end{equation}
with $w_p$ the principal value of the relative tangential velocity on the wing \cite{michelin2008}
\begin{align}
w_p(s_0)=&\Real\Bigg[\ee^{\ci\theta(s_0)}\Bigg(\frac{1}{2\pi\ci}\int_{0}^1\frac{\kk(s)\dd s}{\z(s_0)-\z(s)}+U_\infty-\sum_{j=1}^N\frac{\ci\Gamma_j}{2\pi(\z(s_0)-z_j)}-\dot{\overline{\z}}(s_0)\Bigg)\Bigg].
\end{align}

\subsection{Energy conservation}
From \eqref{eq:beam}, the conservation of energy can be written for the wing
\begin{equation}\label{eq:energy}
\totd{}{t}\left(E_k+E_p\right)=W_p+\mathcal{P}_{in},
\end{equation}
where
\begin{align}
E_k&=\frac{1}{2}\mu\int_0^1|\dot\z|^2\dd s,\\
E_p&=\frac{1}{2}\eta\int_0^1\theta_s^2\dd s,\\
W_p&=-\int_0^1\ddp\Imag\left(\dot\z\ee^{-\ci\theta}\right)\dd s,\\
\mathcal{P}_{in}&=-\Real\left[\left(\dot\z(0)\ee^{-\ci\theta(0)}\right)\left(T(0)+\ci\eta\theta_{ss}(0)\right)\right]-\eta\dot\theta(0)\theta_s(0).
\end{align}
are respectively the kinetic and elastic potential energy of the wing, the rate of work of the pressure forces on the wing and the rate of work of the force and torque applied by the attachment pole on the rest of the wing. In the particular case of a purely heaving motion considered here, $\theta(0)=0$ and $\dot\z(0)=\ci\dot{h}$ is purely imaginary, so
\begin{equation}
\mathcal{P}_{in}=\eta\theta_{ss}(0)\dot{h}.
\end{equation}

\subsection{Propulsive performance}
We are interested in the thrust generated by the flapping wing. The forces applied on the leading-edge attachment are:
\begin{itemize}
\renewcommand{\labelitemi}{$-$}
\item{the elastic forces applied by the sheet on its attachment $\left[T(0)-\ci\eta\theta_{ss}(0)\right]\ee^{\ci\theta(0)}$,}
\item{the force applied by the operator (or animal) to prescribe the leading-edge motion $F_{op}=F_{op}^x+\ci F_{op}^y$,}
\item{the suction force created at the leading edge by the inverse-square root behavior of the pressure. This suction force is the limit of the suction force obtained on a smoothed contour when the curvature radius of the airfoil's leading edge tends to zero. This suction force is equal to \cite{saffman1992,alben2008c,gorelov2008}
\begin{equation}
F_s=-\frac{\pi\ee^{\ci\theta(0)}}{4}\left(\lim_{s\rightarrow 0}\left[\sqrt{s(1-s)}\kappa(s)\right]\right)^2.\end{equation}
}
\end{itemize}
Neglecting the inertia of the attachment and defining the thrust (counted positively to the left) as $\mathcal{T}=-F_{op}^x$, the force balance along the horizontal direction together with \eqref{eq:bc01}--\eqref{eq:bc02} leads to
\begin{equation}
\mathcal{T}=\frac{\pi}{4}\left(\lim_{s\rightarrow 0}\left[\sqrt{s(1-s)}\kappa(s)\right]\right)^2-T(0).
\end{equation}
The instantaneous power input $\mathcal{P}$ by the operator is 
\begin{equation}
\mathcal{P}=\Real\left(F_{op}\dot{\bar{\z_0}}\right)=F_{op}^y\dot{h}(t)=\eta\theta_{ss}(0,t)\dot{h}(t)=\mathcal{P}_{in},
\end{equation}
and is equal to the rate of work $\mathcal{P}_{in}$ of the attachment pole on the wing. Note that this equality would not hold if the motion of the leading edge were a combination of both heaving and pitching, as the suction force $F_s$ would have a non-zero rate of work along the vertical direction.

The useful power output is simply the rate of work  $\mathcal{T}U_\infty$ of the thrust force in the horizontal motion. In non-dimensional units, the flapping efficiency is then defined as the ratio of the average developed thrust to the average input power
\begin{equation}
r=\frac{\langle \mathcal{T}\rangle}{\langle \mathcal{P}^+\rangle},
\end{equation}
where $\langle.\rangle$ is the averaging operator over a flapping period
\begin{equation}
\langle g\rangle=\frac{1}{\tau}\int_0^\tau g(t)\dd t,
\end{equation}
with $\tau=1/\bar{f}$ the non-dimensional period of the flapping motion and $\mathcal{P^+}$ the positive part of $\mathcal{P}$. In the following, the mean power input is understood as $\langle\mathcal{P}^+\rangle$, thereby assuming that the animal cannot store and reuse the energy possibly extracted from the fluid if $\mathcal{P}<0$ during a fraction of the flapping period.  For the range of parameter values used here, it was observed that $\mathcal{P}>0$ for most of the flapping period and $\langle\mathcal{P}^+\rangle\sim\langle\mathcal{P}\rangle$, except for very rigid wings. The results and discussions presented here are therefore not affected by this choice. \\

Finally, the following (non-dimensionalized) quantities are defined for convenience:
\begin{itemize}
\renewcommand{\labelitemi}{$-$}
\item{the trailing-edge peak-to-peak flapping amplitude $\mathcal{D}$,}
\item{the intensity of the wake $\Gamma_m$, defined as the mean value of the amplitude of the successive vortices (positive and negative),}
\item{the induced velocity of the wake vortices $\mathcal{V}$ defined as the horizontal velocity of the wake vortices relative to the imposed unit flow. $\mathcal{V}$ is positive if the wake vortices move faster than the background flow, and negative otherwise.}
\end{itemize}

\section{Numerical simulation of the initial value problem and convergence to a periodic steady state}
\label{sec:numerical}
Equations \eqref{eq:solid1}--\eqref{eq:BM} and \eqref{eq:inteq}--\eqref{eq:bernoulli} are solved numerically expanding $\theta(s,t)$ into a finite series of Chebyshev polynomials of the first kind and using a semi-implicit second-order time-stepping scheme. Taking advantage of the linear relation between $\ddp$ and $\dot\kappa$, added inertia terms can be isolated from the part of the pressure that can be explicitely computed at each time step, thereby avoiding the use of an iterative solver and greatly enhancing the computational efficiency \cite{michelin2008}.

\begin{figure*}
\begin{center}
\includegraphics[width=14.5cm]{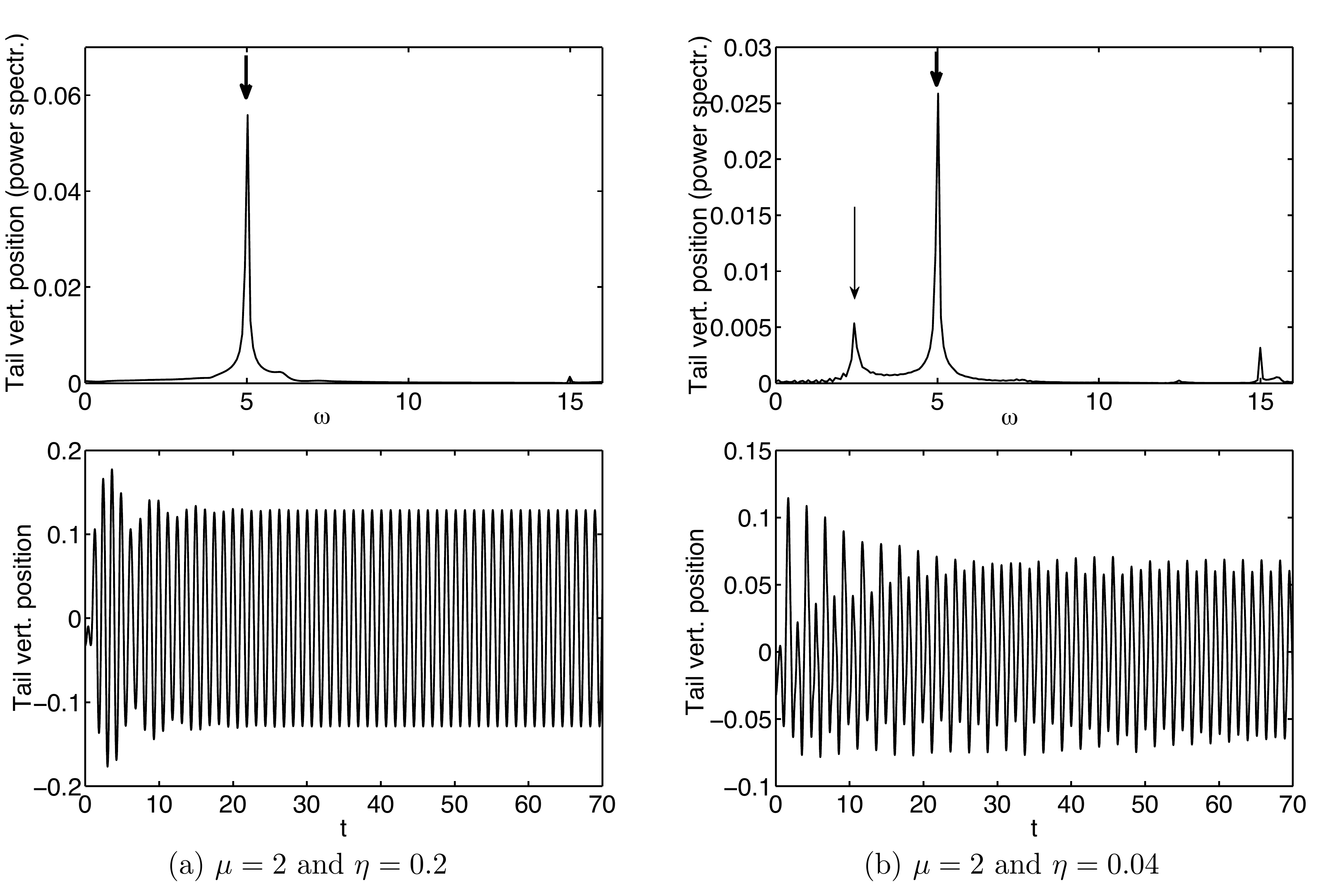}
\caption{(Top) Frequency spectrum and (Bottom) time evolution of the trailing-edge absolute vertical displacement for $\mu=2$, $\varepsilon=0.05$ and $\bar{f}=5/2\pi$. (Left) $\eta=0.2>\eta_m(\mu=2)$ lies in the stability region for the purely passive elastic sheet. The power spectrum displays only one peak (thick arrow) at an angular frequency of $\omega_f=2\pi\bar{f}=5$. (Right) The rigidity $\eta$ is below its critical value $\eta_m(\mu=2)=0.048$ and the elastic sheet is unstable to fluttering. The power spectrum displays two main peaks: one corresponding to the forcing frequency ($\omega_f=5$, thick arrow) and one corresponding to the unstable fluttering mode ($\omega_r\sim 2.3$, thin arrow) that matches the flapping frequency observed in the purely passive case $\varepsilon=0$. In both cases, small peaks can be seen for $\omega\sim 3\omega_f=15$, corresponding to the third harmonic of the forced flapping. The fact that only odd harmonics appear in the tail motion was already previously observed in the case of a passive flag \cite{michelin2008}.}\label{fig:spectra}
\end{center}
\end{figure*}

The system is started from rest. At $t=0$, the horizontal flow is ramped up to its long time unit value and the motion of the leading edge \eqref{eq:heaving} is imposed. After a transient regime of a few heaving periods, a permanent periodic regime is achieved for large enough values of the rigidity $\eta$ (Fig.~\ref{fig:spectra}a). 

However, the harmonic heaving forcing can lead to highly unsteady behaviors if $\eta$ becomes too small: below a certain critical value of the solid's rigidity $\eta_m(\mu)$, the purely passive elastic sheet (as in the flag problem) becomes unstable to fluttering modes and flapping can occur even in the absence of leading-edge forcing \cite{alben2008,michelin2008,alben2008d} ($\eta_m(\mu)$ was found equal to $\eta_m=2\,10^{-3}$ for $\mu=0.2$ and $\eta_m=4.8\,10^{-2}$ for $\mu=2$, using the same point vortex model \cite{michelin2008}). In such cases, the spectrum of the trailing-edge motion can display several peaks, corresponding to the forcing frequency and to the frequency of the unstable modes (Fig.~\ref{fig:spectra}b). The motion is of large enough amplitude for the regime to be non-linear and mode coupling is also expected. As $\eta$ is reduced further, the periodicity is lost and the power spectrum is full; in such a case, determining averaged quantities is not possible anymore.

This explains the difficulty to observe a steady permanent regime when $\eta$ is decreased below the critical value $\eta_m$. This difficulty was not present in the linear study by Alben \cite{alben2008c} as the inertia of the solid was neglected ($\mu\rightarrow 0$). The inertia of the solid is essential to the development of fluttering instability and in the limit $\mu\rightarrow 0$, all the modes are linearly stable \cite{alben2008,alben2008d,michelin2008}. In the following, unless indicated otherwise, the range of parameter values is chosen such that a periodic state is achieved.

\section{Wing flexibility and propulsive performance}
\label{sec:optimal}
In this section, the behavior of the propulsive performance (mean thrust, mean power input and efficiency) is studied when the rigidity $\eta$ of the wing is varied. Several values of the forcing frequency $\bar{f}$, forcing amplitude $\varepsilon$ and mass ratio $\mu$ were investigated.
\subsection{Optimal flexibility for thrust generation and propulsion efficiency}
\label{sec:thrusteff}
\begin{figure*}
\begin{center}
\includegraphics[width=14.5cm]{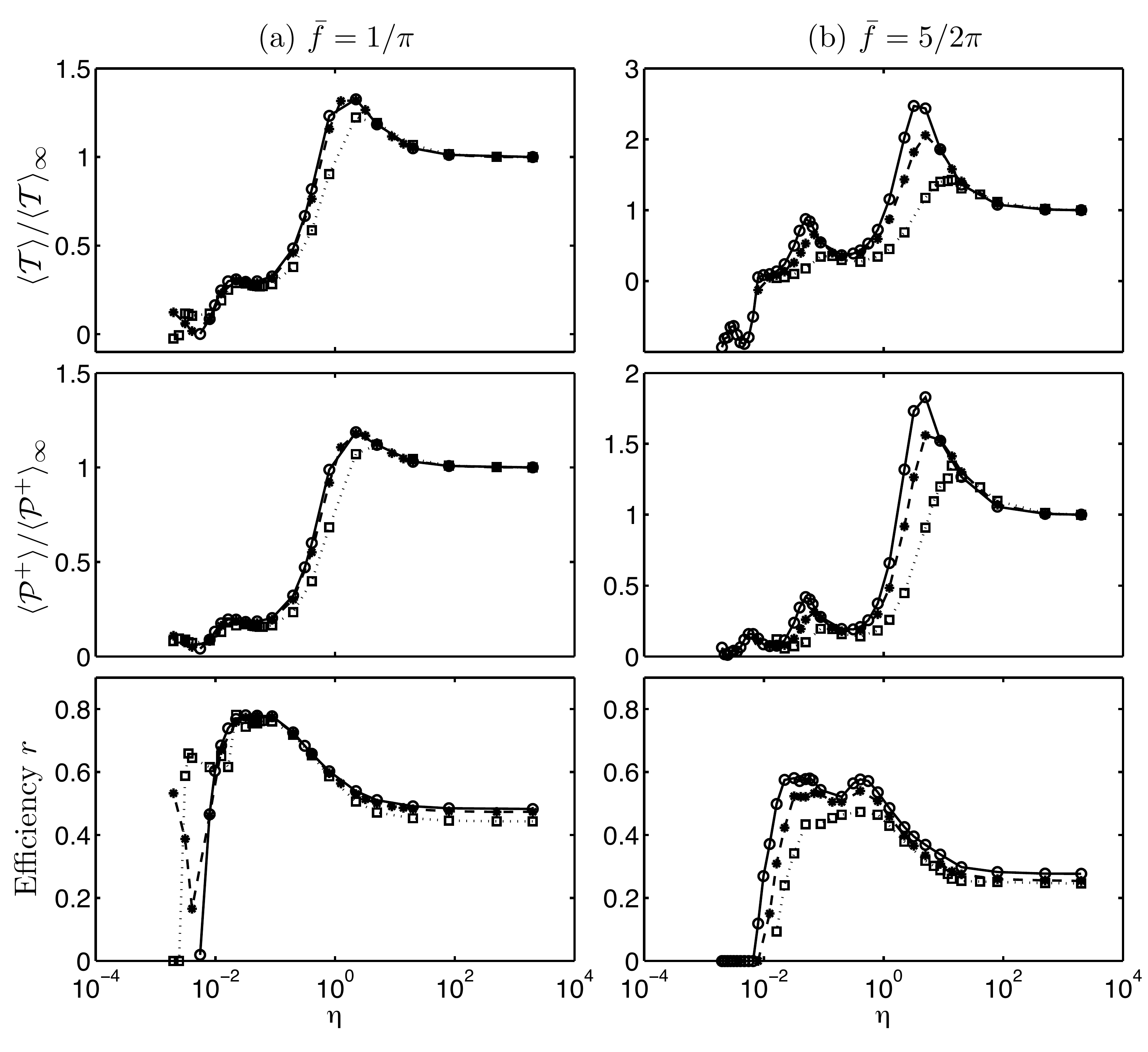}
\caption{Evolution of the mean thrust $\langle \mathcal{T}\rangle$ (top), power input $\langle\mathcal{P}^+\rangle$ (center) and propulsive efficiency (bottom) with the wing's rigidity $\eta$, for a flapping wing of mass ratio $\mu=0.2$, flapping frequency (a) $\bar{f}=1/\pi$ and (b) $\bar{f}=5/2\pi$ and forcing amplitude $\varepsilon=0.1$ (solid-circle), $\varepsilon=0.2$ (dashed-star) and $\varepsilon=0.5$ (dotted-square). For comparison purposes, the mean power input and thrust have been normalized using their rigid case value as $\eta\rightarrow\infty$.}\label{fig:effthrust}
\end{center}
\end{figure*}

For given mass ratio, forcing amplitude and frequency,  the mean thrust, mean power input and propulsive efficiency were computed for each value of the rigidity $\eta$. As a general result, starting from the case of a rigid wing ($\eta\rightarrow\infty$), the mean thrust $\langle \mathcal{T}\rangle$ and power input $\langle\mathcal{P^+}\rangle$ both increase when flexibility is introduced in the problem; however, the former increases faster and the resulting efficiency is an increasing function of flexibility (decreasing function of $\eta$) for large values of $\eta$ (Fig.~\ref{fig:effthrust}). When the solid's flexibility is further reduced, $\langle \mathcal{T}\rangle$ and $\langle\mathcal{P^+}\rangle$ display successive peaks occuring for the same values of $\eta$. The propulsive efficiency $r$ displays in general one large peak (in general for a different value of the rigidity $\eta$ than the thrust peaks) before dropping sharply to zero as the mean thrust vanishes (drag-thrust transition). The increase in the flapping efficiency and developed mean thrust for a flexible wing is significant compared to the case of a rigid wing: for $\bar{f}=5/2\pi$, the peak value of the mean thrust can be greater than twice its value in the rigid case and the efficiency can increase from $27\%$ to almost $60\%$ (Fig.~\ref{fig:effthrust}b). Similar behavior is observed for $\bar{f}=1/\pi$ (Fig.~\ref{fig:effthrust}a). Flexibility has therefore a significant impact on the performance of the propulsive apparatus considered here.

Figure~\ref{fig:effthrust} shows the evolution of the propulsive properties for the lowest possible value of $\eta$ leading to a permanent periodic regime. In the higher frequency case ($\bar{f}=5/2\pi$), the drag-thrust transition was clearly observed while at lower frequency ($\bar{f}=1/\pi$), this transition is less well defined as unsteady phenomena start developing around the same value of the rigidity. Several factors lead to unsteadiness of the problem, including transitions in the wake behind the flapping wing and development of the fluttering instability for low values of the rigidity (see section \ref{sec:numerical}).

The behavior of $\langle\mathcal{T}\rangle$, $\langle\mathcal{P}^+\rangle$ and $r$ collapse rather well for leading-edge flapping amplitudes up to $50\%$ of the wing's length and lower frequencies (Fig.~\ref{fig:effthrust}a). However, for higher frequencies, the influence of the forcing amplitude can be seen for values of $\varepsilon$ as low as $0.1$ (Fig.~\ref{fig:effthrust}b). As a general result, the propulsive efficiency and the normalized mean thrust and power input are observed to decrease with $\varepsilon$ when all other parameters are held fixed. The optimal values of $\eta$ for maximal thrust or maximal efficiency are also observed to increase with $\varepsilon$.

It must be emphasized here that the decrease of the achievable thrust and power input with the forcing amplitude $\varepsilon$ are only relative to the rigid case: for higher forcing amplitudes, the absolute mean thrust and power input are larger in magnitude and, in the rigid limit, are observed to scale like $\varepsilon^{5/2}$.

\subsection{Wake structure and thrust production}
\begin{figure}
\begin{center}
\includegraphics[width=8.6cm]{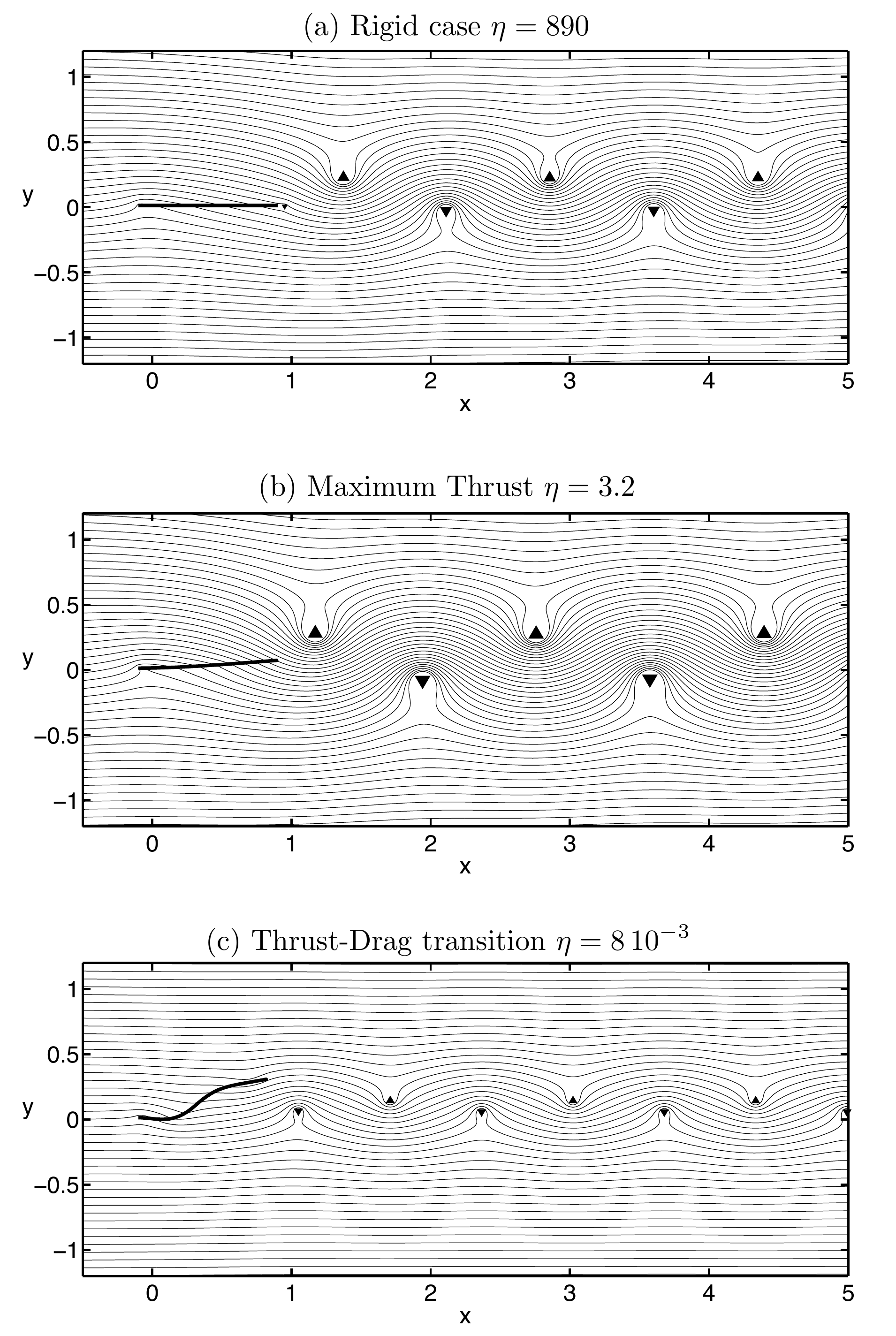}
\caption{Streamlines of the flow over the flapping wing for $\mu=0.2$, $\varepsilon=0.1$ and $\bar{f}=5/2\pi$ and decreasing rigidity $\eta$. The positive (resp. negative) vortices are represented with upward- (resp. downward-) pointing triangles with sizes scaled to the intensity of the vortices. The streamlines are plotted for $t=20$ when the leading edge crosses the horizontal axis, and the leading-edge forcing for $0\leq t\leq 20$ was the same in all three cases.}\label{fig:strlines}
\end{center}
\end{figure}

\subsubsection{Evolution of the wake structure with increasing flexibility}
For given mass ratio, frequency and forcing amplitude $\varepsilon$, the variation of the flexibility of the wing induces important changes in the developed thrust and energy use. As $\eta$ is decreased from the rigid case ($\eta\rightarrow\infty$), the wake behind the flapping wing also undergoes important modifications. Figure~\ref{fig:strlines} shows the evolution of the flow pattern around the heaving wing for varying $\eta$. In the high-rigidity case (Fig.~\ref{fig:strlines}a), the trailing-edge deflection is small. The wake vortices are arranged in a reverse Von K\'arm\'an vortex street, in which vortices with positive (resp. negative) intensity are positioned above (resp. below) the horizontal axis. This arrangement induces an acceleration of the fluid on the horizontal axis to the right and the vortices have a higher velocity than the imposed background flow (positive $\mathcal{V}$). This jet carries momentum to the right which is consistent with the formation of a mean thrust on the wing. The formation of a reversed Von K\'arm\'an vortex street in thrust-producing propulsion schemes is well known and has been observed in several experimental studies \cite{anderson1998,godoydiana2008}. 

When $\eta$ is decreased, the solid is more flexible and the trailing-edge flapping amplitude $\mathcal{D}$ increases. As a result, the intensity of the wake vortices increases with decreasing rigidity and so does the thrust generated by the flapping motion. Figure~\ref{fig:strlines}(b) shows the streamlines and arrangement of the wake vortices for $\eta=3.2$ which corresponds to the first peak in thrust production on Fig.~\ref{fig:effthrust}(b). The phase between the leading and trailing-edge displacements has also been modified compared to the rigid case. The increase in the vortex advection velocity can be seen as the distance between two successive vortices is slightly increased (the vortex shedding frequency is unchanged and equal to the forcing frequency). 

As a comparison, Fig.~\ref{fig:strlines}(c) shows the flow pattern for the value of $\eta$ corresponding to the thrust-drag transition. One observes immediately that the intensity of the wake vortices has decreased significantly and the width of the vortex street has almost vanished. In intermediate $\Rey$ experiments, the formation of a classical Von-K\'arm\'an street is associated with drag on the generating solid body. It appears here that, even at the transition between thrust and drag production, the reversed Von K\'arm\'an pattern persists, although largely weakened. This result is consistent with the observation in recent experiments \cite{godoydiana2008} that the thrust-drag transition does not necessarily occur at the same time as the transition in the wake structure. A theoretical argument for this difference is also presented in section \ref{sec:lambsaff}. Furthermore, in purely inviscid simulations, the Von K\'arm\'an street in the drag-producing case is generally much weaker and instead the vortices tend to align along the axis \cite{alben2008,michelin2008}.

\subsubsection{Classical and reversed Von K\'arm\'an streets and thrust/drag production}
\label{sec:lambsaff}
Here, theoretical results on vortex streets are used to understand the relationship between drag/thrust production and vortex wake structures. We focus on highly periodic cases where the structure of the reversed Von-K\'arm\'an or classical Von K\'arm\'an streets is easily identified. 

\begin{figure}
\begin{center}
\includegraphics[width=8.6cm]{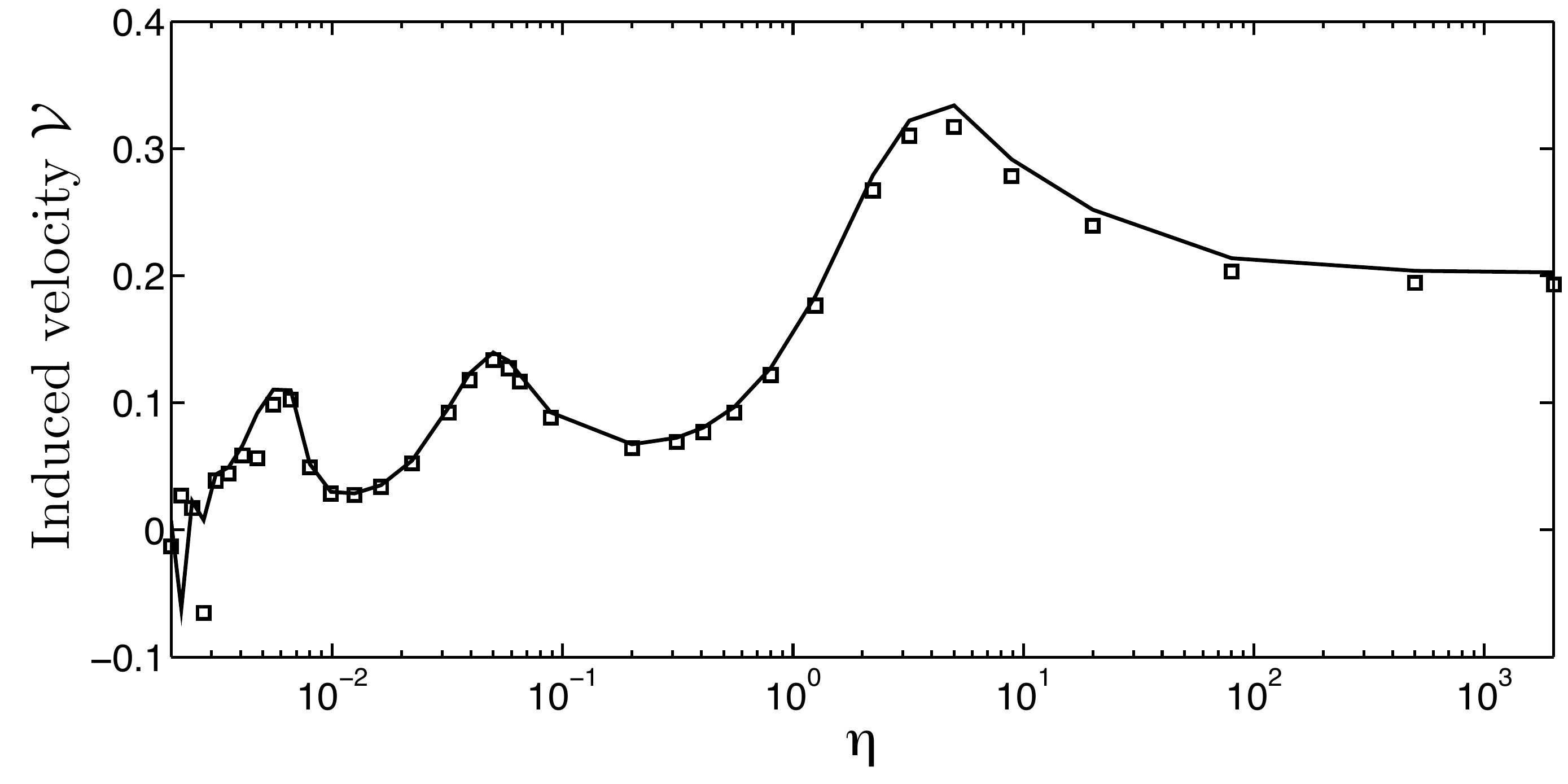}
\caption{(Square) Induced wake vortices velocity $\mathcal{V}$ behind the flapping wing for varying $\eta$ and $\mu=0.2$, $\varepsilon=0.1$ and $\bar{f}=5/2\pi$. The predicted value obtained for an infinite reversed Von-K\'arm\'an vortex street \cite{lamb1932} is plotted for comparison (solid line).}\label{fig:lamb}
\end{center}
\end{figure}

For a given vortex wake intensity and vortex arrangement, the advection velocity of the vortices is the superposition of the incoming flow velocity (equal to $1$ in non-dimensional units) and of the induced velocity $\mathcal{V}$ of the vortex street, which is itself a direct function of the width of the vortex street (vertical distance between the two rows of opposite sign vortices), its intensity and its wavelength (horizontal distance between two successive vortices of identical sign). If the vortex street were infinite, its induced velocity $\mathcal{V}$ would be \cite{lamb1932}
\begin{equation}
\label{eq:lamb1}
\mathcal{V}=\frac{\Gamma_w}{2a}\tanh\left(\frac{\pi b}{a}\right),
\end{equation}
where $\Gamma_w$ is the magnitude of the vortices, $b$ the width of the vortex street and $a$ its horizontal wavelength. In \eqref{eq:lamb1}, the following convention is chosen: $b>0$ (resp. $b<0$) for a reversed (resp. classical) Von-K\'arm\'an street in which positive  vortices are located above (resp. below) the horizontal axis. If the shedding frequency (equal to the flapping frequency  $\bar{f}$) is fixed, then $a=(1+\mathcal{V})/\bar{f}$. Measuring the width of the vortex street $b$ and its intensity $\Gamma_w$, \eqref{eq:lamb2} leads to a non-linear equation for $\mathcal{V}_L$, the predicted value of the induced velocity, that can be solved numerically for given $b$ and $\Gamma_w$
\begin{equation}
\label{eq:lamb2}
\mathcal{V}_L=\frac{\Gamma_w\bar{f}}{2(1+\mathcal{V}_L)}\tanh\left(\frac{\pi b\bar{f}}{(1+\mathcal{V}_L)}\right).
\end{equation}
Figure~\ref{fig:lamb} shows very good agreement between the measured value $\mathcal{V}$ and the expected value $\mathcal{V}_L$ for varying $\eta$ (the other parameters taking the same values as in Fig.~\ref{fig:strlines}), particularly above the thrust-drag transition at $\eta=8\,10^{-3}$ (Fig.~\ref{fig:lamb}). This agreement shows that the induced velocity is mostly determined by the neighboring vortices, and the semi-infinite or infinite nature of the vortex street does not influence the induced velocity significantly.

\begin{figure}
\begin{center}
\includegraphics[width=8.6cm]{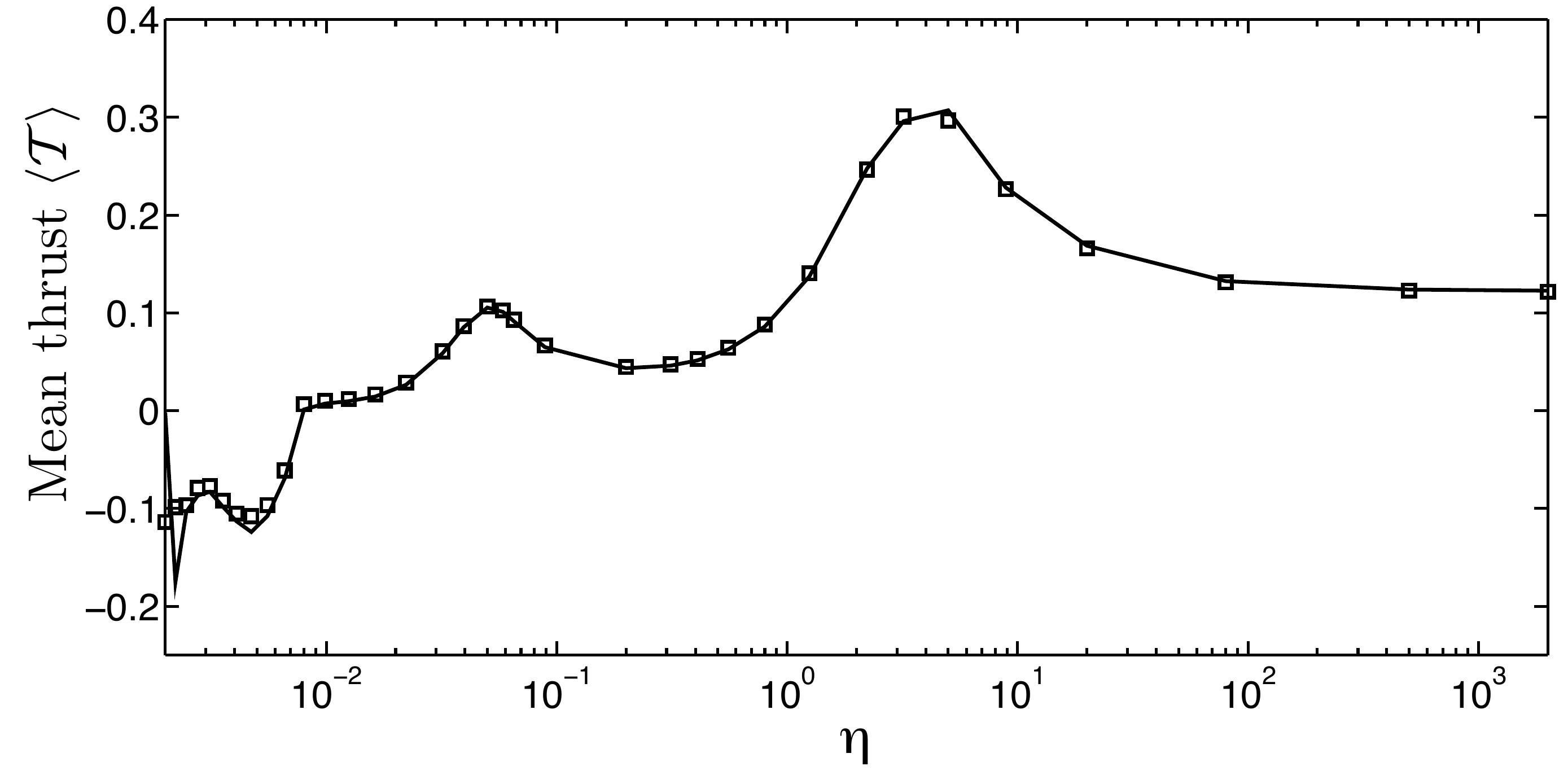}
\caption{(Square) Mean thrust $\langle\mathcal{T}\rangle$ produced by the flapping wing for varying $\eta$ and $\mu=0.2$, $\varepsilon=0.1$ and $\bar{f}=5/2\pi$. The predicted value $\mathcal{T}_S$ \eqref{eq:saff} based on impulse conservation \cite{kochin1964,saffman1992} is also plotted for comparison (solid line). $\mathcal{T}_S$ was computed from \eqref{eq:saff} using the values of the vortex street intensity, width and induced velocity obtained with our model.}\label{fig:saffman}
\end{center}
\end{figure}

The drag on a solid body placed in a uniform parallel flow and shedding vortices in the pattern of a staggered vortex street (regular or reversed Von-K\'arm\'an street) was computed by Von K\'arm\'an using the conservation of momentum around the solid body and part of its wake \cite{kochin1964, saffman1992}. The only important elements in the derivation are again the intensity $\Gamma_w$, wavelength $a$ and width $b$ of the vortex street. In particular, the motion (other than the main translation) and deformation of the body are irrelevant in Von K\'arm\'an's derivation. These results can easily be generalized to the case of a reversed Von-K\'arm\'an street and in our notation, the predicted mean thrust $\mathcal{T}_S$ is obtained as
\begin{equation}
\label{eq:saff}
\mathcal{T}_S=-\frac{\Gamma_w^2}{2\pi a}+\frac{\Gamma_w\,b}{a}\left(1+2\mathcal{V}\right),\quad \textrm{with   }a=(1+\mathcal{V})/\bar{f}.
\end{equation}
This theoretical prediction compares very well to the results obtained for the mean thrust in our simulations (Fig.~\ref{fig:saffman}). In particular, the transition between thrust and drag production is well reproduced. The agreement is lost at low values of $\eta$: For such low values of the rigidity, the highly regular structure of the wake is also lost as some natural modes of the passive elastic sheet become unstable to fluttering and the motion of the wing loses its strong periodicity, making the definition of $a$, $\Gamma_w$ and $b$ difficult.

In \eqref{eq:saff}, the thrust consists of two terms. The second term is positive in the case of a reversed Von K\'arm\'an street ($b>0$) and negative in the classical Von K\'arm\'an street (with $b<0$ and $\mathcal{V}<0$, as long as $\mathcal{V}>-1/2$, which always occur in the weak Von K\'arm\'an streets observed here). The presence of the first term (which is always negative and therefore always leads to drag production) is responsible for the experimentally- and numerically-observed difference between the thrust/drag transition and the transition in the wake structure. As in the passive flapping flag case \cite{alben2008,michelin2008}, it is possible to have net drag produced by a weak reversed Von K\'arm\'an street as the first term in \eqref{eq:saff} dominates the second one.

\subsection{Resonance and optimal flexibility for thrust production}
\begin{figure}
\begin{center}
\includegraphics[width=8.6cm]{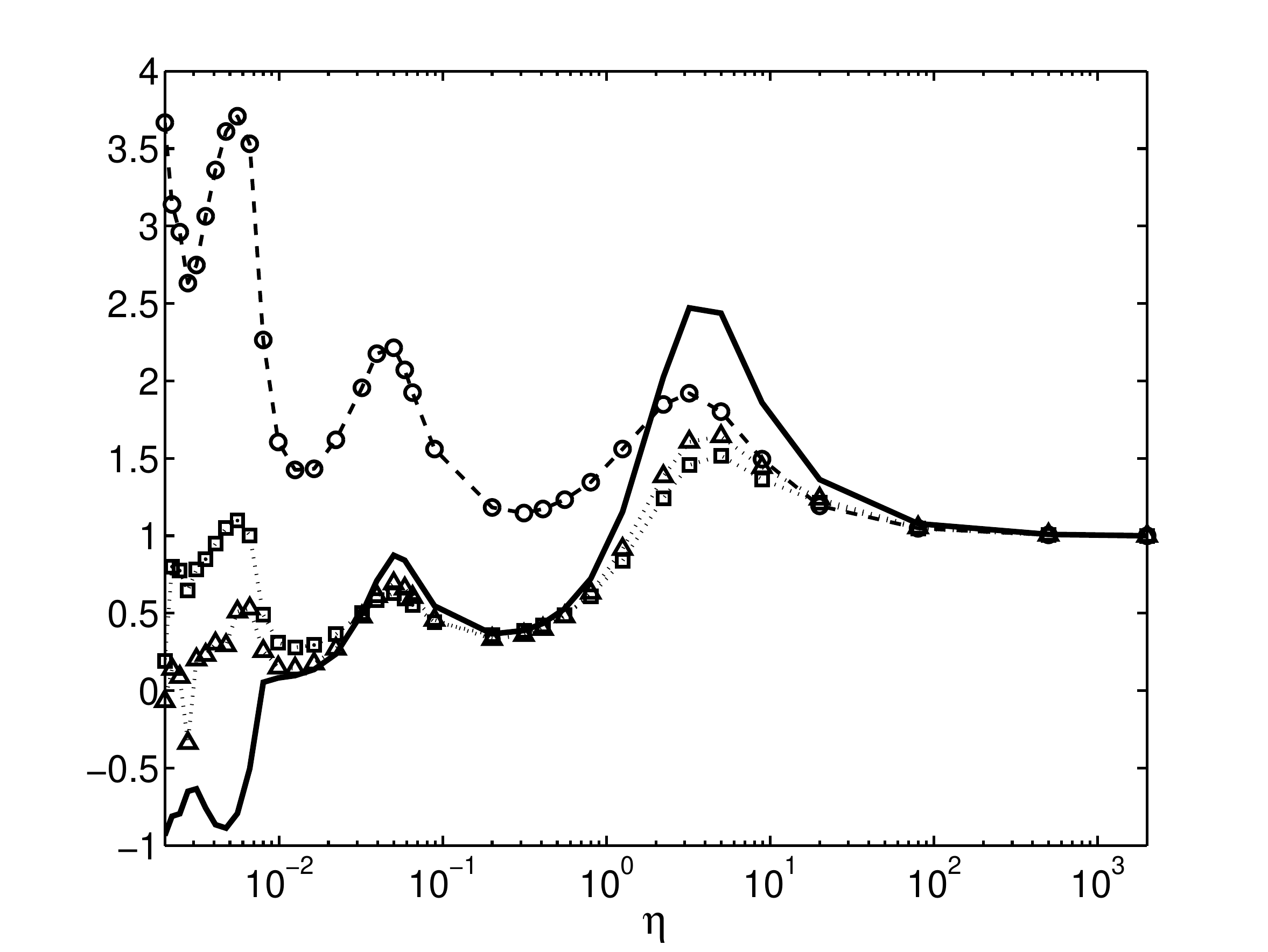}
\caption{Evolution of the mean thrust $\langle\mathcal{T}\rangle$ (solid), trailing-edge flapping amplitude $\mathcal{D}$ (dashed-circle), vortex wake intensity $\Gamma_m$ (dotted-square) and induced vortex velocity $\mathcal{V}$ (dotted-triangle) for $\mu=0.2$, $\varepsilon=0.1$ and $\bar{f}=5/2\pi$.  All quantities have been normalized by their rigid-case value ($\eta\rightarrow\infty$).}\label{fig:res1}
\end{center}
\end{figure}

In Fig.~\ref{fig:effthrust}(b), successive peaks in the mean thrust created by the heaving wing can be observed, but also a peak in the mean drag (or negative thrust) below the drag-thrust transition that occurs around $\eta\sim 8\,.\,10^{-3}$.  Figure~\ref{fig:res1} shows the evolution of the mean thrust, flapping amplitude $\mathcal{D}$, wake intensity $\Gamma_w$ and wake induced velocity $\mathcal{V}$ for the same values of the parameters $\mu$, $\varepsilon$ and $\bar{f}$ as in Fig.~\ref{fig:effthrust}.  

A very clear correlation is observed between the occurence of the maxima (in magnitude) for the mean thrust (or drag) and for the other quantities, which confirms the argument presented in the previous section: the mean thrust and drag peaks are created by an increase in the flapping motion at the trailing edge, where the vortex wake is formed. An increase in $\mathcal{D}$ (with a constant flapping frequency) induces higher relative velocity at the trailing edge, and therefore stronger shed vortices. While this increase in $\Gamma_m$ with $\mathcal{D}$ is physical, it is however not possible to find a simple scaling of $\Gamma_m$ with $\mathcal{D}$, as other factors that depend on $\eta$ must be taken into account (e.g. the orientation of the trailing edge relative to its velocity). The induced velocity on the vortex street $\mathcal{V}$ is therefore also increased and the wake carries a larger fluid momentum downstream, thereby creating a greater thrust on the profile.

It is also interesting to notice that the maximum drag observed for $\eta\sim 5.\,10^{-3}$ is also associated with a maximum in $\mathcal{D}$ and $\Gamma_m$. In that case, the large amplitude of motion at the trailing edge opposes the imposed flow and creates a net drag on the body.

\section{Influence of flexibility on the flapping amplitude and mode shape}
\label{sec:resonance}

\subsection{Resonances between the forcing frequency and the natural frequencies of the system}
In this section, we are interested in the origin of the maxima in the trailing-edge flapping amplitude $\mathcal{D}$. The successive peaks in $\mathcal{D}$ as $\eta$ is varied suggest a resonance phenomenon. By varying the rigidity of the solid $\eta$, the natural frequencies of the system are also modified. For an elastic sheet in vacuum, these frequencies scale like $\sqrt{B/\rho_s}$. Looking for solutions of \eqref{eq:solid1}--\eqref{eq:bc2} in the linear limit with no fluid forcing, the fundamental angular frequencies  $\omega_{0n}=2\pi f_n$ of a clamped-free elastic sheet in vacuum are obtained in non-dimensional form as
\begin{equation}
\omega_{0n}=\lambda^2_n\sqrt{\frac{\eta}{\mu}},\qquad \textrm{with   } 1+\cosh\lambda_n\cos\lambda_n=0.
\end{equation}
The ratio $\sqrt{\mu/\eta}$ can be thought of as the non-dimensional time-scale associated with the frequency of the sheet's natural oscillations in vacuum, or alternatively as the outside flow velocity non-dimensionalized by the characteristic velocity associated with the sheet's properties.

\subsubsection{Influence of $\mu$ and $\varepsilon$ on the position of the resonance peaks in $\mathcal{D}$}
\begin{figure}
\begin{center}
\includegraphics[width=8.6cm]{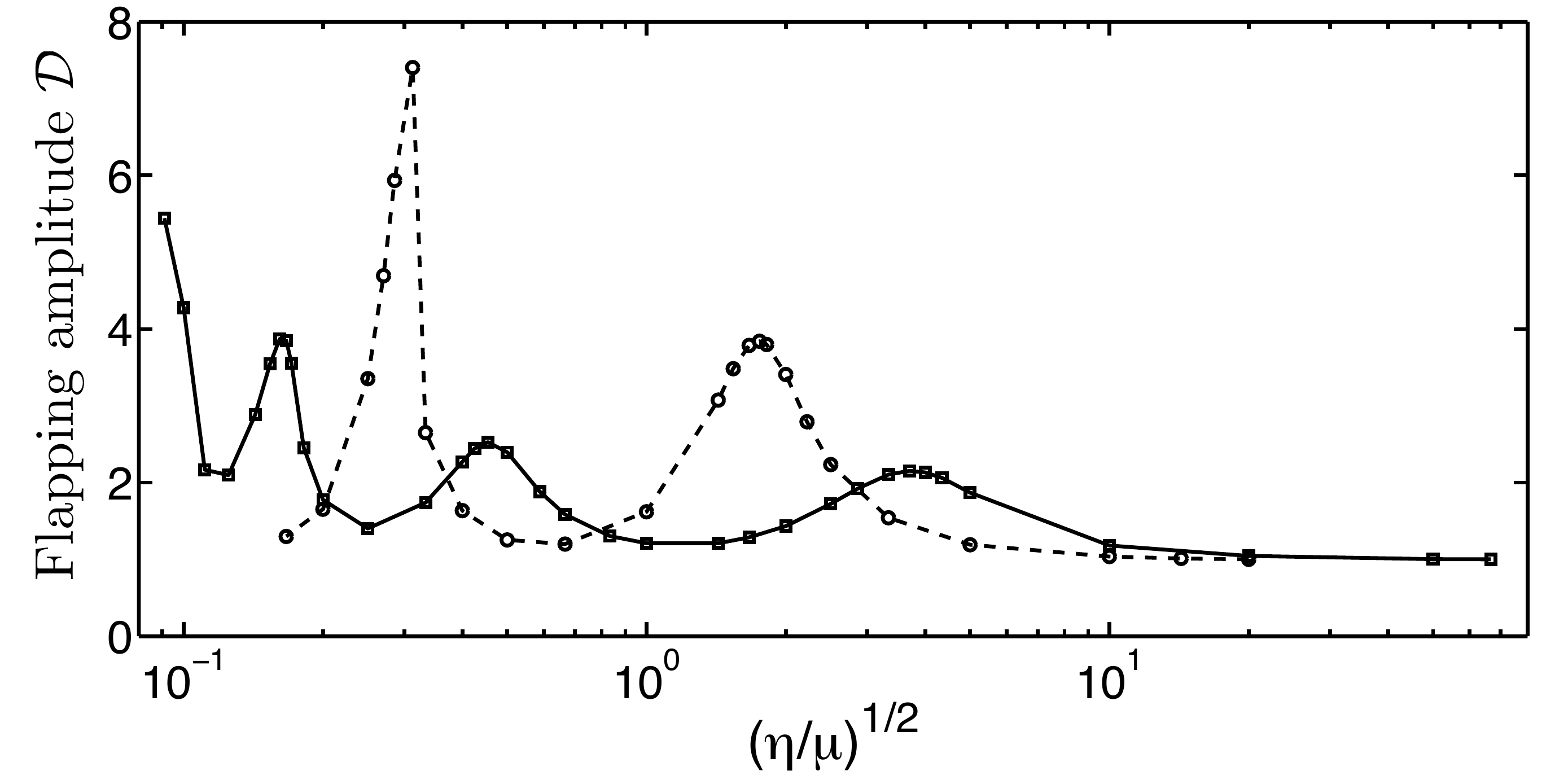}
\caption{Evolution of the normalized trailing-edge flapping amplitude $\mathcal{D}$ with $\sqrt{\eta/\mu}$ for $\varepsilon=0.05$, $\bar{f}=5/2\pi$ and for $\mu=0.2$ (solid-squares) and $\mu=2$ (dashed-circle). The flapping amplitude was normalized using the asymptotic rigid case limit ($\eta\rightarrow \infty$).}\label{fig:compmu}
\end{center}
\end{figure}

However, the resonance peaks observed in Fig.~\ref{fig:res1} do not correspond to a resonance with the natural frequency of the elastic sheet in vacuum. If this were the case, then the resonance would be achieved for all values of the mass ratio $\mu$ at the same value of $\sqrt{\eta/\mu}$. Such a coincidence does not occur (Fig.~\ref{fig:compmu}): the position of the successive resonances is actually strongly influenced by the fluid-solid inertia ratio $\mu$. This difference makes sense physically, as the eigenfrequencies of the system are modified by the presence of the forcing horizontal flow, and such effects as added inertia are expected to be important. Instead, the natural frequencies of the system \{wing + outside uniform flow\} should be considered. A comparison with linear analysis prediction is proposed below in section \ref{sec:complin}.

\begin{figure}
\begin{center}
\includegraphics[width=8.6cm]{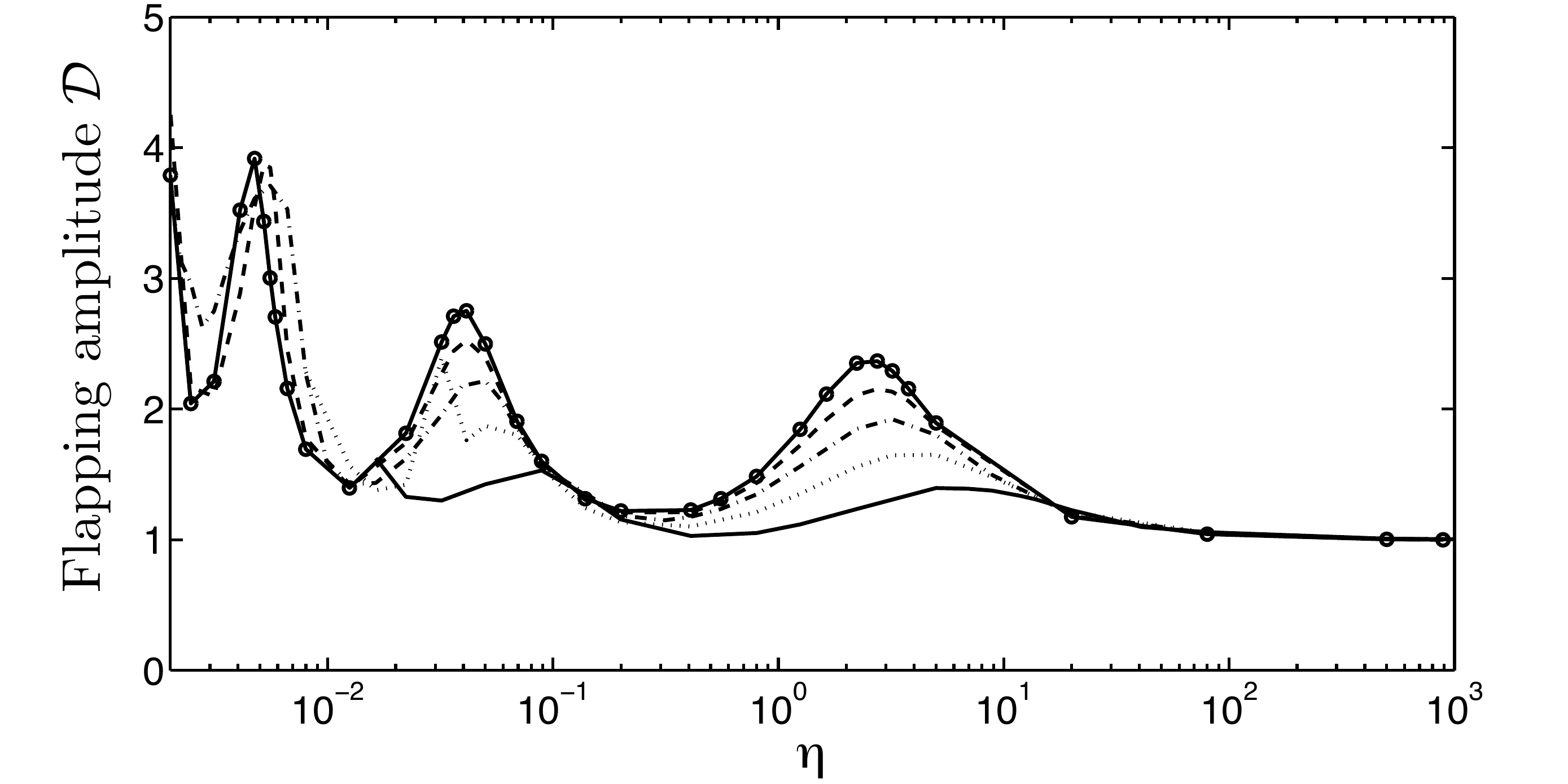}
\caption{Influence of the forcing amplitude $\varepsilon$ on the position of the resonances in the trailing-edge flapping amplitude for $\mu=0.2$ and $\bar{f}=5/2\pi$. The results are plotted for $\varepsilon=0.01$ (solid-circles), $\varepsilon=0.05$ (dashed), $\varepsilon=0.1$ (dash-dotted), $\varepsilon=0.2$ (dotted) and $\varepsilon=0.5$ (solid). The trailing-edge flapping amplitude was normalized by its rigid-case value ($\eta\rightarrow\infty$).}\label{fig:compeps}
\end{center}
\end{figure}

Before comparing the numerical results to the linear analysis, we study the influence of the forcing amplitude $\varepsilon$ on the position of the resonances. It was noticed in section \ref{sec:thrusteff} that the normalized thrust and power input follow similar patterns for values of $\varepsilon$ up to $0.2$ to $0.5$ depending on the value of $\mu$. In an attempt to determine the exact value of $\eta$ leading to a resonance, we study how the resonance peaks in $\mathcal{D}$ are modified as $\varepsilon$ is varied between $0.01$ and $0.5$. Starting from small $\varepsilon$, one observes that the increase in $\varepsilon$ induces a shift of the resonance peaks toward larger values of $\eta$ and a smoothing of the resonance peaks (Fig.~\ref{fig:compeps}). Also, the convergence toward the limit case of small $\varepsilon$ is faster for the second and third resonance peaks than for the first one (labeling the peaks from the right as $\eta$ is decreased from the rigid case limit). 

\subsubsection{Absolute and relative trailing-edge flapping amplitude}
\begin{figure}
\begin{center}
\includegraphics[width=8.6cm]{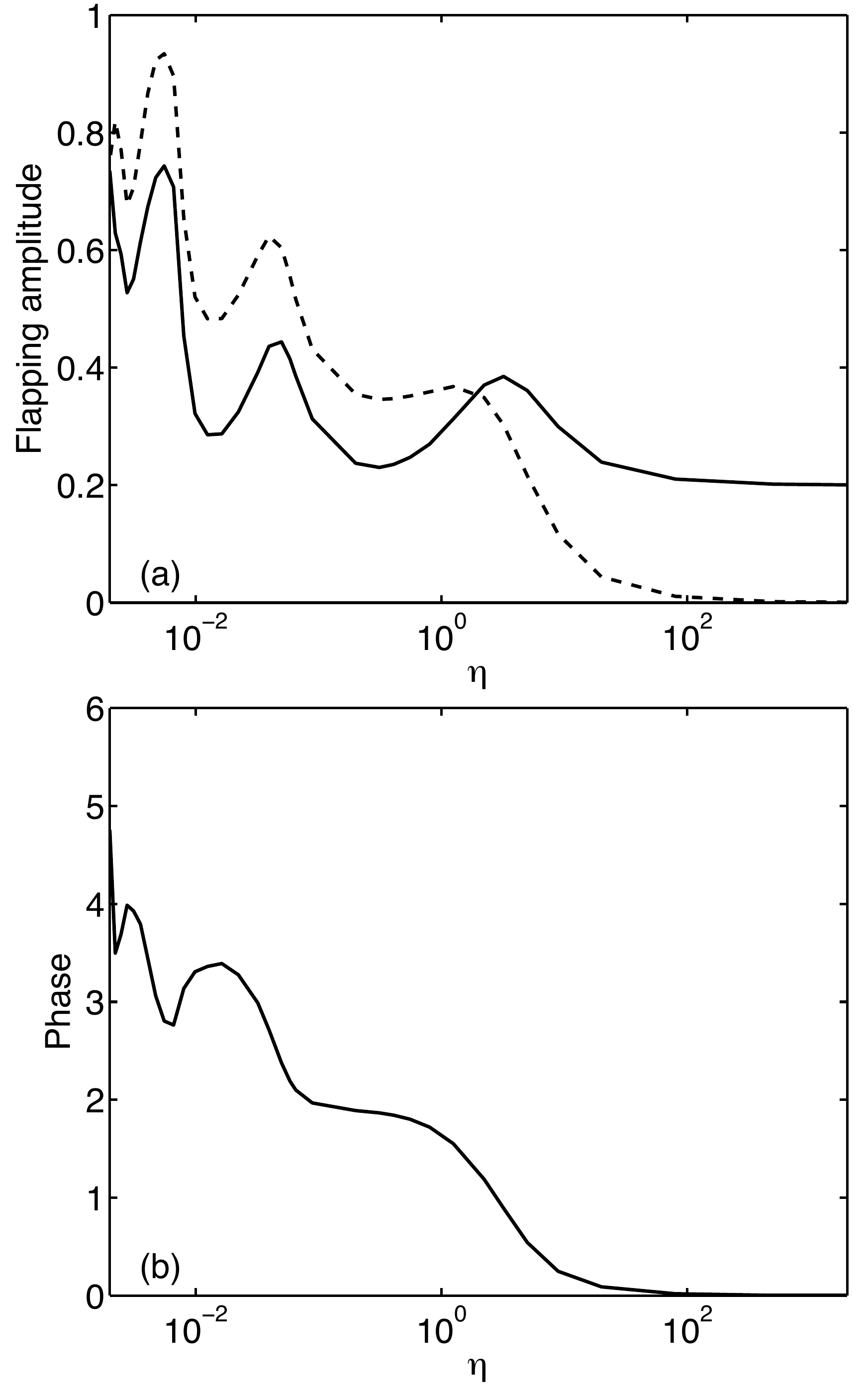}
\caption{(a) Comparison between the absolute flapping amplitude $\mathcal{D}$ (solid) and relative flapping amplitude $\mathcal{D}^*$ (dashed) for the trailing edge, for $\mu=0.2$, $\bar{f}=5/2\pi$ and $\epsilon=0.1$. (b) Phase difference between the leading-edge and trailing-edge motions.}\label{fig:phase}
\end{center}
\end{figure}

In the previous sections, $\mathcal{D}$ was defined as the peak-to-peak amplitude of the trailing edge flapping motion in the laboratory frame (thereafter referred to as absolute flapping amplitude). In the following, we are also interested in the motion of the wing \emph{in the frame moving with the leading edge}. In this frame, the peak-to-peak trailing edge flapping amplitude is $\mathcal{D}^*$ (thereafter referred to as relative flapping amplitude). $\mathcal{D}$ and $\mathcal{D}^*$ are not necessarily equal because of the phase difference between the motion of the leading and trailing edges (Fig.~\ref{fig:phase}). This delay is, in general, a decreasing function of the solid's rigidity. In the limit of a rigid wing $\eta\rightarrow\infty$, the trailing edge flaps in phase with the leading edge with the same amplitude ($\mathcal{D}^*=0$ and $\mathcal{D}=2\varepsilon$). As $\eta$ is decreased, a delay appears between the motion of the leading and trailing edges: the elasticity of the wing takes more time to carry along its length the signal imposed at the leading edge. As a result, the relative amplitude $\mathcal{D}^*$ increases.

One of the main consequences of the appearance of such a delay is the non-coincidence of the first resonance peak (with largest $\eta$) in the flapping amplitude, whether $\mathcal{D}$ or $\mathcal{D}^*$ is considered. The position of the subsequent peaks are not significantly affected (Fig.~\ref{fig:phase}a)

\subsubsection{Comparison with the natural frequencies of the system}
\label{sec:complin}
In this section, the conjecture that the trailing-edge flapping amplitude peaks are due to a resonance with the natural frequencies of the \{wing + imposed flow\} system is tested by comparing the value of the rigidity $\eta$ leading to a peak value for $\mathcal{D}$ to the value of $\eta$ for which the forcing frequency matches a natural frequency of the system. The position of the peaks in $\mathcal{D}$ is measured in the limit of small forcing amplitude (typically $\varepsilon=0.01$). The natural frequencies of a purely passive wing clamped at its leading edge in a uniform flow are determined using the linear stability analysis method developed by Kornecki \cite{kornecki1976}. A brief summary of this method is given in the Appendix \ref{sec:linstab}. For given $\mu$ and $\eta$, the eigenfrequencies of the system are computed. The lowest frequencies are also associated with the modes of lowest order (those with the longest wavelength). Here, the following equivalent problem is considered: for a given $\mu$ and $\bar{f}$, we want to find the values $\eta$ for which a mode of the passive elastic sheet in axial flow has the particular frequency $\bar{f}$, regardless of its growth rate.

\begin{figure}
\begin{center}
\includegraphics[width=8.6cm]{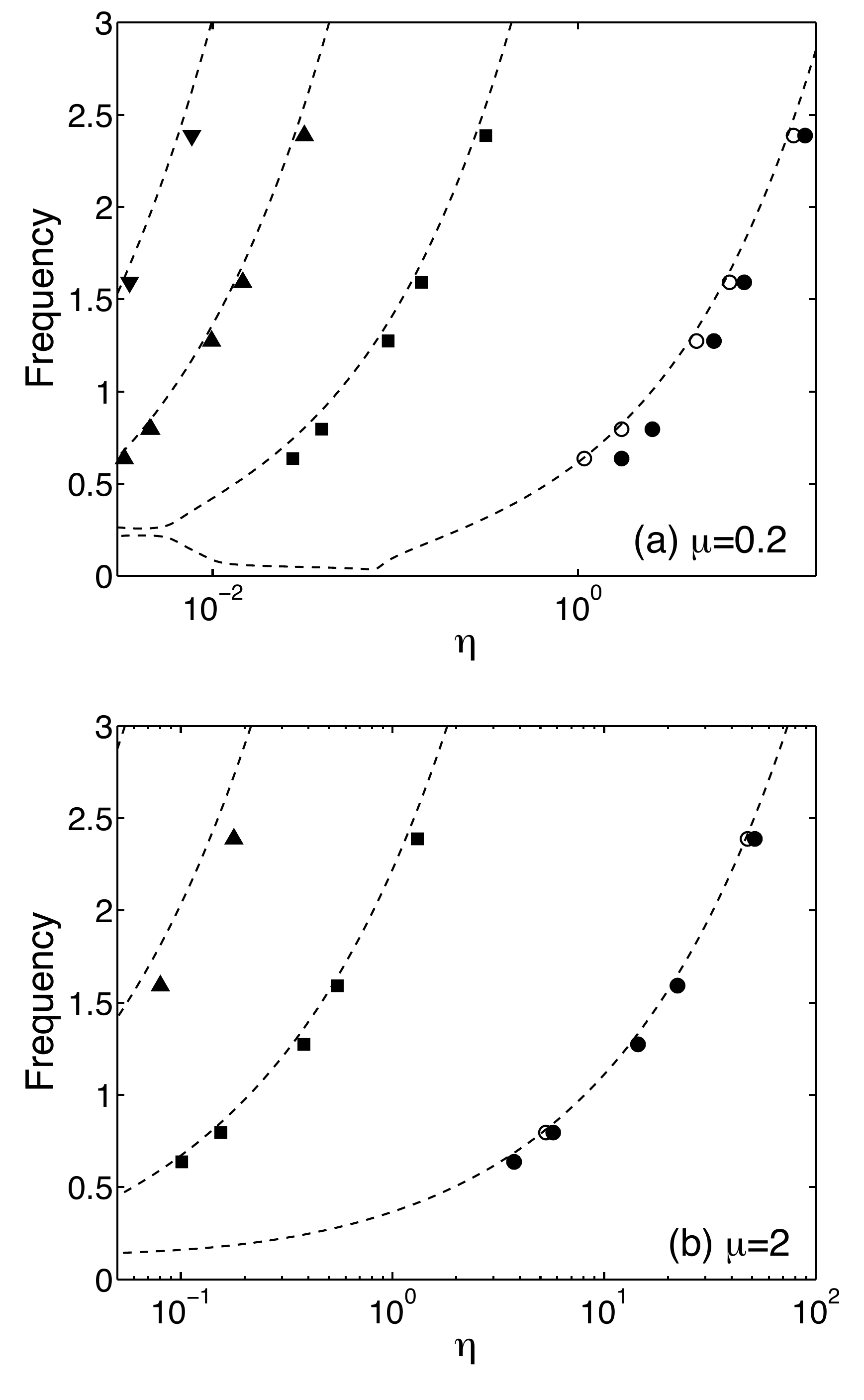}
\caption{Position of the resonances for the trailing-edge flapping amplitude in the $(\eta,\bar{f})$-plane obtained using the present model (symbols) with $\varepsilon=0.01$ and (a) $\mu=0.2$ and (b) $\mu=2$. The different symbols correspond to the nature of the mode: modes $1$ have no neck in the flapping enveloppe (circles); modes $2$ (squares), modes $3$ (upward-pointing triangle) and modes $4$ (downward-pointing triangles) have respectively $1$, $2$ and $3$ necks in their motion envelope. The black symbols correspond to resonances in the absolute flapping amplitude $\mathcal{D}$. Open symbols correspond to resonances in the relative flapping amplitude $\mathcal{D}^*$ when they differ from the resonances in $\mathcal{D}$. The position of the resonances is compared to the prediction of the linear analysis (dashed) for the natural frequency of the purely passive elastic sheet (or flag) in axial flow (see Appendix \ref{sec:linstab}).}\label{fig:kornecki}
\end{center}
\end{figure}

Figure~\ref{fig:kornecki} shows the position in the $(\eta,\bar{f})$-plane of the first resonances observed in the trailing-edge flapping amplitude $\mathcal{D}$ for small heaving amplitude $\varepsilon$, starting from the rigid case $\eta\rightarrow\infty$. Although $\varepsilon$ is small, $\mathcal{D}$ can be significant (greather than $20\varepsilon$ at the resonance in the case of the heavier wing ($\mu=2$), thereby representing more than $20\%$ of the wing's length). These results are compared to the natural frequencies of the system \{wing + parallel flow\} as predicted by the linear analysis and very good agreement is found between the numerical results and the theoretical predictions. The position of the resonances in $\mathcal{D}^*$ are also indicated. Resonances in $\mathcal{D}^*$ and $\mathcal{D}$ coincide except for the first resonance peak in the case of the smaller mass ratio $\mu$.

The linear analysis seems to underpredict slightly the values of $\eta$ corresponding to the resonance for a given frequency $\omega$. This difference is consistent with the amplitude of discrepancy between the point vortex model and the linear stability analysis observed in the study of a purely passive elastic sheet or flag \cite{michelin2008}. Two other factors can also explain the small discrepancy in the results:
\begin{itemize}
\renewcommand{\labelitemi}{$-$}
\item{As pointed out, the motion of the wing is not infinitesimally small here, even for $\varepsilon=0.01$; it is therefore possible that non-linear effects modify the exact position of the resonances. In the previous section, the effect of increasing $\varepsilon$ was shown to shift the resonance peaks toward larger $\eta$, particularly for the first peak, and this could account for a significant part of the observed discrepancy.}
\item{The amplitude of flapping in the laboratory frame is considered here. However, even for small $\varepsilon$, this amplitude differs from the relative flapping amplitude defined in the frame moving with the leading edge (see Fig.~\ref{fig:phase}), because of the existence of a non-zero phase between the motion of the leading and trailing edges. It was observed in Fig.~\ref{fig:phase} that the resonance in relative amplitude occurs for smaller values of $\eta$, particularly for the first resonance (largest $\eta$). In the linear analysis, both amplitudes (relative and absolute) are identical since the leading edge is held fixed. For $\mu=0.2$, the agreement is improved for the position of the first resonance if $\mathcal{D}^*$ is considered instead of $\mathcal{D}$ (Fig.~\ref{fig:kornecki}).}
\end{itemize}

Figure~\ref{fig:kornecki} also indicates the nature of the observed mode, in particular the number of necks in the envelope of the wing's motion. One observes that resonances located on a same branch of the linear analysis prediction share the same general structure, and the number of necks is consistent with that predicted by the linear analysis: for given $\mu$ and $\eta$, the lowest frequency mode has the longest wavelength and no neck in its envelope. The next lowest frequency corresponds to a mode with one neck, and so on for the successive frequencies. The evolution of the mode shape for varying $\eta$ is studied in more detail in section \ref{sec:modeshape}.

\subsection{Evolution of the flapping mode shape with the flexibility of the profile}
\label{sec:modeshape}
\begin{figure*}
\begin{center}
\includegraphics[width=15cm]{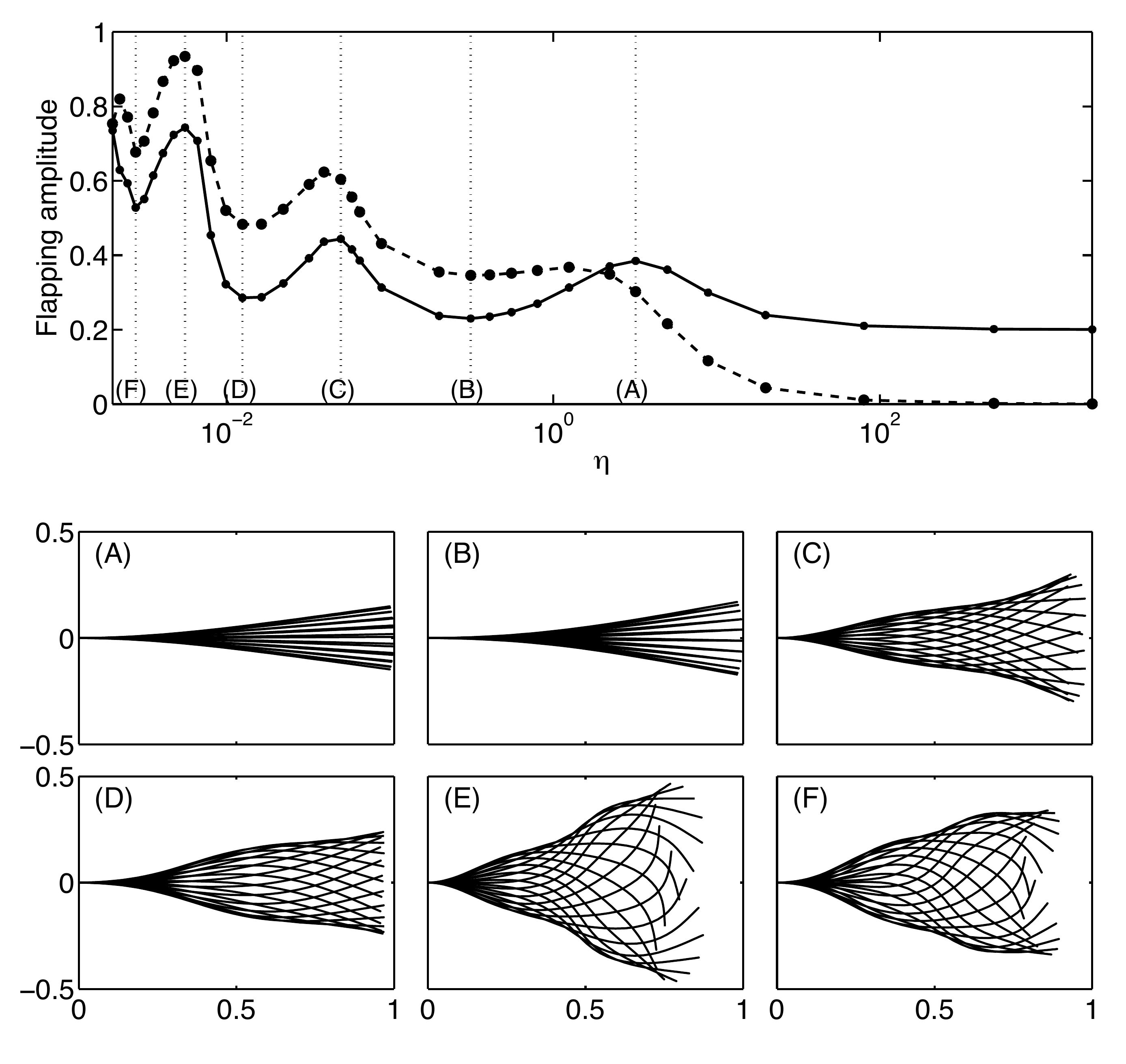}
\caption{Evolution of the mode shape with $\eta$ for $\mu=0.2$, $\varepsilon=0.1$ and $\bar{f}=5/2\pi$. (Top) Trailing-edge flapping amplitude in the stationary frame $\mathcal{D}$ (solid) and in the frame attached to the leading edge $\mathcal{D}^*$ (dotted). Note that the value of $\eta$ leading to a resonance in $\mathcal{D}$ does not necessarily correspond to the value of $\eta$ leading to a resonance in $\mathcal{D}^*$. (Bottom) Mode shape plotted for the value of $\eta$ indicated on the top panel. The position of the wing in the frame moving with the leadinge edge is plotted every $\Delta t=0.06$.}\label{fig:modeshape}
\end{center}
\end{figure*}
To confirm that the maxima of flapping amplitude actually correspond to resonances between the forcing frequency and the natural frequency of the passive elastic sheet in a parallel flow, the evolution of the flapping mode shape with $\eta$ is now considered. For comparison with the case of a purely passive flexible elastic sheet, the mode shape is defined as the envelope of the motion of the wing \emph{in the frame moving with the leading edge}. Figure~\ref{fig:modeshape} shows the mode shape in permanent regime at the values of $\eta$ leading to peak values of the absolute trailing-edge flapping amplitude and to the minima between two successive peaks. Note that on Fig.~\ref{fig:modeshape}, case $B$ seems to correspond to a wider envelope than case $A$, although the absolute flapping amplitude is smaller for $B$ than for $A$. This is the result of the change of frame: if one considers the flapping amplitude in the moving frame, the position of the maxima differs from peaks in absolute flapping amplitude (see Fig.~\ref{fig:phase}). Cases $A$, $C$ and $E$ correspond to the resonances while $B$, $D$ and $F$ correspond to local minima of the absolute flapping amplitude.

The mode shapes $C$ and $E$ are structurally similar to the envelope of the first two flapping modes observed for the passive flexible flag \cite{michelin2008}, which is consistent with $C$ and $E$ corresponding to resonances between the flapping frequency and the flapping modes $2$ and $3$ (one- and two-neck modes respectively). $A$ corresponds to mode $1$ of the passively flapping elastic sheet, which was not observed in the case of the flapping flag study \cite{michelin2008} as it is always stable \cite{eloy2007} and therefore does not lead to spontaneous large-scale flapping.

\section{Conclusions}\label{sec:conclusions}
Using a reduced-order model for the flow past a two-dimensional heaving flexible wing, the influence of the wing's flexibility on its propulsive performance (mean thrust, flapping efficiency) was investigated. Starting from the purely rigid case, we observed that the flexibility of the wing allowed for a larger trailing-edge flapping amplitude, thereby generating a stronger wake and an increased mean thrust. The energy usage also increases with the introduction of flexibility, but more slowly than the mean thrust, resulting in a net increase in the flapping efficiency with reduced rigidity. This efficiency gain can be significant (up to twice the efficiency of the rigid case). While the mean thrust and power input display several peaks when the rigidity $\eta$ is decreased from the purely rigid case, the flapping efficiency displays one wide peak before falling sharply as $\eta$ nears the value leading to the thrust-drag transition. Below this threshold, the wing is too flexible to communicate momentum to the flow and instead starts creating a net drag on the leading-edge attachment.

The relationship between thrust production and wake structure was then investigated, taking advantage of the discrete representation of the wake. Analytical predictions for the induced vortex street velocity and mean thrust in terms of the vortex street strength and spatial arrangement were successfully compared to our simulation results.

The peaks in mean thrust were found to correspond to maximum values in the trailing-edge amplitude, and shown to be the result of the resonance between the forcing frequency of the heaving motion and the natural frequencies of the system. A quantitative comparison showed very good agreement between the optimal values of the solid's rigidity and the linear analysis predictions for the resonances position. The existence of these resonance phenomena was further confirmed by comparing the flapping mode shape to the mode shape observed for a freely flapping elastic sheet (e.g.~the flag problem\cite{michelin2008}).

The natural frequencies of the system are strongly dependent on both the solid's flexibility and the ratio of fluid and solid inertia, and so are the optimal values of the solid's rigidity. For a slender neutrally buoyant fish fin, the inertia ratio $\mu$ can generally be neglected \cite{alben2008c}. However, in the case of an insect wing, the small thickness-to-chord ratio is balanced by the large difference in density for the fluid and solid, and the present analysis shows that the mass ratio $\mu$ plays an important role in defining the optimal value of the flexibility.

The model presented here used a potential flow representation and the shedding of point vortices to describe the highly-unsteady flow around the insect wing. Vortices were shed only from the trailing edge. This approximation is reasonable for small angles of attack (proportional to the Strouhal number $\mbox{\textit{St}}$ in the case of a purely heaving motion). Then, the vorticity shed at the leading edge is negligible or merges with the vorticity shed by separation of the boundary layers at the trailing edge \cite{anderson1998}, leading to the shedding of two individual vortices every flapping period and so-called $2S$ wakes \cite{williamson1988}. However, when the flapping amplitude and frequency are increased, more complex wakes  are expected as vorticity is shed from both the trailing and leading edges, inducing for example the formation of the so-called $2P$ wake, where a vortex pair is shed during each half period.  The representation of the leading-edge vortex falls however beyond the scope of inviscid methods such as the present point vortex model or a vortex sheet approach as the effect of viscosity can not be neglected \cite{michelin2009a,michelin2009b}: the leading-edge vortex is expected to remain close to the body for a sufficiently long time to interact with the boundary layers. The use of point vortices also restricts this method to two-dimensional problems and therefore does not allow the study of such effects as wing-tip vortices that are expected to influence significantly the flight performance. The present study however does not aim at reproducing the exact flow around an insect wing but to present a test case where the effect of flexibility can be isolated from other factors such as leading-edge and wing-tip vortices.

Despite these limitations, the present method offers the advantage of a considerable reduction in computational cost for two-dimensional fluid-solid simulations. This is particularly attractive for situations where the cost of full numerical simulation is prohibitive and a large number of simulations are required (typically in the case of optimization problems).

The present work was purposely limited to a one-degree of freedom flapping pattern (pure heaving) in order to limit the number of free parameters and focus on the fundamental effect of solid flexibility on the flapping performance. In future work, more realistic flapping schemes should be considered to follow more closely the flapping pattern of an insect wing. In particular, a combination of heaving and pitching should be considered to determine the influence of the relative phase between heaving and pitching motions on the results presented in this work. The interaction between two flapping flexible sheets should also be investigated to understand the lift and thrust generation by insects with multiple pairs of flexible wings (e.g. dragonfly \cite{wang2007}) or the efficiency of fish schooling, and to complement recent experimental studies on multiple passive flexible filaments \cite{zhang2000,jia2007,jia2008,ristroph2008}.

\section*{Acknowledgments}
This work was supported by the Human Frontier Science Program Research Grant RGY 0073/2005.

\appendix*

\section{Natural frequencies of a passive flexible sheet in axial flow}
\label{sec:linstab}
We present here a brief summary of the linear stability analysis of a clamped-free flexible sheet or flag using a vortex sheet representation of the wake as developed by Kornecki \cite{kornecki1976}. More details on the method and the full calculation can be found in Refs. \cite{bisplinghoff1955,kornecki1976}. Small vertical displacements of the sheet $h(s,t)$ are considered ($0\leq s\leq 1$) with $|h|\ll 1$ so that the equation of motion of the solid \eqref{eq:beam} becomes in linearized form
\begin{equation}
\mu\ddot{h}=-\eta h_{ssss}-\Delta p.
\end{equation}
The wake is represented by a continuous distribution of vorticity $\gamma(s,t)$ along the horizontal axis ($s\geq 1$) advected by the flow (this continuous shedding thereby differs from the discrete approach presented in the present paper, but is better suited to linear stability analysis). The pressure forcing can be decomposed into two parts \cite{theodorsen1935,bisplinghoff1955,kornecki1976}: a non-circulatory part due to the flow created by the solid's motion with no net circulation around the solid and a circulatory part due to the flow created by the vortex wake (the absence of circulation at infinity requiring the existence of a net circulation around the solid equal to the opposite of the wake vorticity). Considering a decomposition of the solid's motion onto normal modes of the form $h(s,t)=\Real\left[y(s)\ee^{\ci\omega t}\right]$, $y(s,t)$ satisfies \cite{kornecki1976}
\begin{widetext}\begin{align}\label{eq:theod}
-\mu\omega^2y+\eta y_{ssss}=&\,-\frac{2\ci\omega}{\pi}\int_0^1\log\left|\frac{\sqrt{\frac{x}{1-x}}+\sqrt{\frac{\xi}{1-\xi}}}{\sqrt{\frac{x}{1-x}}-\sqrt{\frac{\xi}{1-\xi}}}\right|\left(\ci\omega y+y_s\right)\dd\xi\nonumber\\
&+\frac{2}{\pi\sqrt{x(1-x)}}\int_0^1\frac{\sqrt{\xi(1-\xi)}}{x-\xi}\left(\ci\omega y+y'\right)\dd\xi\\
&-\frac{2}{\pi}\int_0^1\sqrt{\frac{\xi}{1-\xi}}\left(\ci\omega y+y_s\right)\dd\xi\left[\frac{2x-1}{\sqrt{x(1-x)}}+\sqrt{\frac{1-x}{x}}\,C(\omega)\right],\nonumber\\
\end{align}
\end{widetext}
where $C(\omega)$ is the Theodorssen function \cite{bisplinghoff1955}:
\begin{equation}
C(\omega)=\frac{2H^{(2)}_1\left(\frac{\omega}{2}\right)}{H^{(2)}_1\left(\frac{\omega}{2}\right)+\ci H^{(2)}_0\left(\frac{\omega}{2}\right)},
\end{equation}
and $H^{(2)}_\nu(x)=J_\nu(x)-\ci Y_\nu(x)$ ($\nu=0,1$) are Hankel functions of the second kind \cite{abramowitz1964}.
 
\begin{figure*}
\begin{center}
 \includegraphics[width=16.5cm]{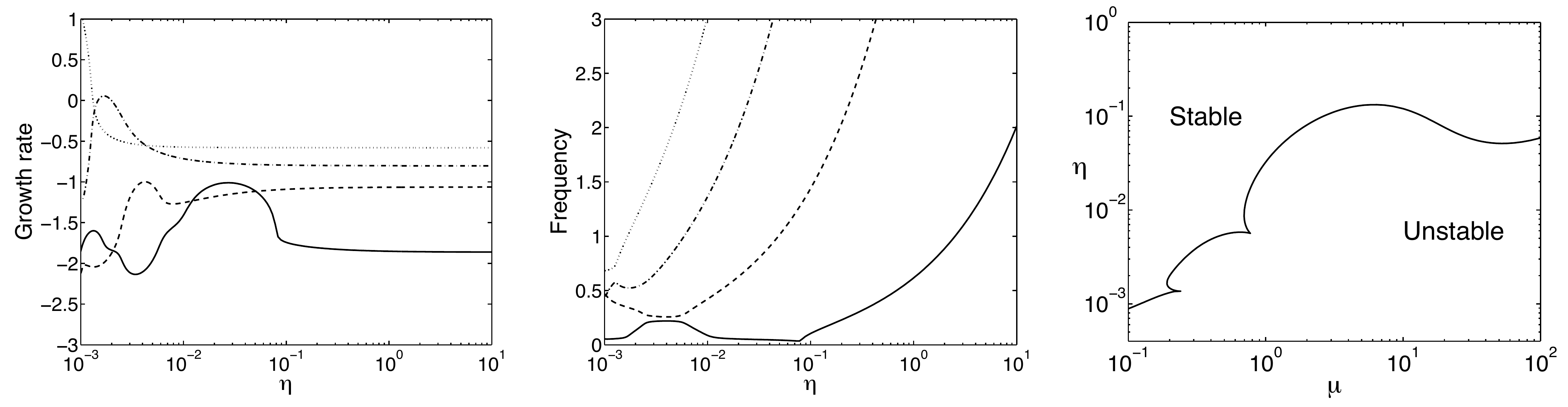}
 \caption{For $\mu=0.2$, evolution of (Left) the growth rate and (Center) frequency of the first four modes (of lowest positive frequency): mode $1$ (solid), mode $2$ (dashed), mode $3$ (dash-dotted) and mode $4$ (dotted). (Right) Critical stability curve as obtained in the ($\mu$,$\eta$)-plane from the linear stability analysis. }\label{fig:korneckires}
 \end{center}
 \end{figure*}

The eigenvalue problem \eqref{eq:theod} for $\omega$ is solved numerically using a Galerkin method: $y(s)$ is decomposed along the first $N$ eigenmodes of the clamped-free beam in vacuum $y(s)=\sum\alpha_n\psi^{(n)}(s)$ with $\psi^{(n)}(s)$ satisfying
\begin{align*}
\psi^{(n)}_{ssss}&=\lambda_n^4\psi^{(n)},\\
\psi^{(n)}(0)=\psi^{(n)}_s(0)&=\psi^{(n)}_{ss}(1)=\psi^{(n)}_{sss}(1)=0,
\end{align*}
and $\lambda_n$ are the successive positive roots of $1+\cos\lambda_n\cosh\lambda_n=0$.
\eqref{eq:theod} is replaced by the nonlinear eigenvalue problem 
\begin{align}
\Bigg[-\omega^2\left(\mu\mathbf{I}+ \mathbf{M^{(M)}}\right)&+\ci\omega \mathbf{M^{(G)}}+\left(\eta \mathbf{K}+\mathbf{M^{(K)}}\right)+C(\omega)\mathbf{M^{C1}}+\ci\omega C(\omega)\mathbf{M^{C2}}\Bigg].\bm{\alpha}=0.\label{eq:eigdis}
\end{align}
The coefficients of the different $N\times N$ matrices are defined as
\begin{align*}
K_{ij}&=\lambda_j^4\delta_{ij},\\
M^{(K)}_{ij}&=\mathcal{F}_1(\psi^{(i)},\psi^{(j)}_x)+\mathcal{F}_2(\psi^{(i)},\psi^{(j)}_x),\\
M^{(G)}_{ij}&=\mathcal{F}_1(\psi^{(i)},\psi^{(j)})+\mathcal{F}_2(\psi^{(i)},\psi^{(j)})+\mathcal{F}_3(\psi^{(i)},\psi^{(j)}_x), \\
M^{(M)}_{ij}&=\mathcal{F}_3(\psi^{(i)},\psi^{(j)})\\
M^{C1}_{ij}&=\mathcal{F}_4(\psi^{(i)},\psi^{(j)}_x), \\
M^{C2}_{ij}&=\mathcal{F}_4(\psi^{(i)},\psi^{(j)}),
\end{align*}
with the functionals $\mathcal{F}_k$ ($1\leq k\leq 4$) defined as
\begin{align*}
\mathcal{F}_1(f,g)&=-\frac{2}{\pi}\int_0^1\frac{f(x)}{\sqrt{x(1-x)}}\left[\int_0^1\frac{\sqrt{\xi(1-\xi)}}{x-\xi}g(\xi)\dd\xi\right]\dd x\\
\mathcal{F}_2(f,g)&=\frac{2}{\pi}\left[\int_0^1\frac{(2x-1)f(x)}{\sqrt{x(1-x)}}\dd x\right]\left[\int_0^1\sqrt{\frac{x}{1-x}}g(x)\dd x\right],\\
\mathcal{F}_3(f,g)&=\frac{2}{\pi}\int_0^1f(x)\left[\int_0^1g(\xi)\log\left|\frac{\sqrt{\frac{x}{1-x}}+\sqrt{\frac{\xi}{1-\xi}}}{\sqrt{\frac{x}{1-x}}-\sqrt{\frac{\xi}{1-\xi}}}\right|\dd\xi\right]\dd x,\\
\mathcal{F}_4(f,g)&=\frac{2}{\pi}\left[\int_0^1f(x)\sqrt{\frac{1-x}{x}}\dd x\right]\left[\int_0^1\sqrt{\frac{x}{1-x}}g(x)\dd x\right].
\end{align*}
For a given $N$, these matrices can be precomputed. Then, for given $\mu$ and $\eta$, the eigenvalue problem \eqref{eq:eigdis} is solved iteratively using a Newton-Kantorovitch algorithm \cite{boyd2001}. Figure~\ref{fig:korneckires} shows the evolution of the growth rate $-\Imag(\omega)$ and frequency $\omega/2\pi$ for $\mu=0.2$ and varying rigidity $\eta$. In the rigid case (large $\eta$), all modes are stable. As $\eta$ is decreased below $\eta_m=2.\,10^{-3}$, one or more modes become unstable (mode $3$ then mode $4$). The critical stability curve separates the region of the ($\mu$,$\eta$)-plane where the elastic sheet's state of rest is stable and the region of the parameter space where at least one mode is unstable (Fig.~\ref{fig:korneckires}).

\begin{thebibliography}{10}

\bibitem{lighthill1960}
M.~J. Lighthill.
\newblock Note on the swimming of slender fish.
\newblock {\em J. Fluid Mech.}, 9:305--317, 1960.

\bibitem{lighthill1969}
M.~J. Lighthill.
\newblock Hydromechanics of aquatic animal propulsion.
\newblock {\em Ann. Rev. Fluid Mech.}, 1:413--446, 1969.

\bibitem{childress1981}
S.~Childress.
\newblock {\em Mechanics of Swimming and Flying}.
\newblock Cambridge University Press, Cambridge, 1981.

\bibitem{triantafyllou2000}
M.~S. Triantafyllou, G.~S Triantafyllou, and D.~K.~P. Yue.
\newblock Hydrodynamics of fishlike swimming.
\newblock {\em Annu. Rev. Fluid Mech.}, 32:33--53, 2000.

\bibitem{wang2005}
Z.~J. Wang.
\newblock Dissecting insect flight.
\newblock {\em Annu. Rev. Fluid Mech.}, 37:183--210, 2005.

\bibitem{thomas2004}
A.~L.~R. Thomas, G.~K. Taylor, R.~B. Srygley, R.~L. Nudds, and R.~J. Bomphrey.
\newblock Dragonfly flight: free-flight and tethered flow visualizations reveal
  a diverse array of unsteady lift-generating mechanisms, controlled primarily
  via angle of attack.
\newblock {\em J. Exp. Biol.}, 207:4299--4323, 2004.

\bibitem{ellington1984}
C.~P. Ellington.
\newblock The aerodynamics of hovering insect flight.
\newblock {\em Phil. Trans. R. Soc. London \emph{B}}, 305:1--181, 1984.

\bibitem{dickinson1999}
M.H. Dickinson, F.~O. Lehmann, and S.~P. Sane.
\newblock Wing rotation and the aerodynamic basis of insect flight.
\newblock {\em Science}, 284:1954--1960, 1999.

\bibitem{wang2004b}
Z.~J. Wang, J.~M. Birch, and M.~H. Dickinson.
\newblock Unsteady forces and flows in low reynolds number hovering flight:
  two-dimensional computations vs robotic wing experiments.
\newblock {\em J. Exp. Biol.}, 207:449--460, 2004.

\bibitem{pesavento2004}
U.~Pesavento and Z.~J. Wang.
\newblock Falling paper: {N}avier--{S}tokes solutions, model of fluid forces,
  and center of mass elevation.
\newblock {\em Phys. Rev. Lett.}, 93:144501, 2004.

\bibitem{berman2007}
G.~Berman and Z.~J. Wang.
\newblock Energy-minimizing kinematics in hovering insect flight.
\newblock {\em J. Fluid Mech.}, 582:153--168, 2007.

\bibitem{anderson1998}
J.~M. Anderson, K.~Streitlien, D.~S. Barrett, and M.~S. Triantafyllou.
\newblock Oscillating foils of high propulsive efficiency.
\newblock {\em J. Fluid Mech.}, 360:41--72, 1998.

\bibitem{godoydiana2008}
R.~Godoy-Diana, J.~L. Aider, and J.~E. Wesfreid.
\newblock Transitions in the wake of a flapping foil.
\newblock {\em Phys. Rev. E}, 77:016308, 2008.

\bibitem{wu1961}
T.~Y. Wu.
\newblock Swimming of a waving plate.
\newblock {\em J. Fluid Mech.}, 10:321--344, 1961.

\bibitem{wang2000a}
Z.~J. Wang.
\newblock Two dimensional mechanism for insect hovering.
\newblock {\em Phys. Rev. Lett.}, 85:2216--2219, 2000.

\bibitem{wang2000b}
Z.~J. Wang.
\newblock Vortex shedding and frequency selection in flapping flight.
\newblock {\em Phys. Rev. Lett.}, 410:323--341, 2000.

\bibitem{miao2006}
J.-M. Miao and M.-H. Ho.
\newblock Effect of flexure on aerodynamic propulsive efficiency of flapping
  flexible airfoil.
\newblock {\em J. Fluids Struct.}, 22:401--419, 2006.

\bibitem{shukla2007}
R.~K. Shukla and J.~D. Eldredge.
\newblock An inviscid model for vortex shedding from a deforming body.
\newblock {\em Theor. Comput. Fluid Dyn.}, 21:343--368, 2007.

\bibitem{wootton1992}
R.~J. Wooton.
\newblock Functional morphology of insect wings.
\newblock {\em Ann. Rev. Entomol.}, 37:113--140, 1992.

\bibitem{combes2003a}
S.~A. Combes and T.~L. Daniel.
\newblock Flexural stiffness in insect wings {I}. {S}caling and the influence
  of wing venation.
\newblock {\em J. Exp. Biol.}, 206:2979--2987, 2003.

\bibitem{combes2003b}
S.~A. Combes and T.~L. Daniel.
\newblock Flexural stiffness in insect wings {II}. {S}patial distribution and
  dynamic wing bending.
\newblock {\em J. Exp. Biol.}, 206:2989--2997, 2003.

\bibitem{connell2007}
B.~S.~H. Connell and D.~K.~P. Yue.
\newblock Flapping dynamics of a flag in uniform stream.
\newblock {\em J. Fluid Mech.}, 581:33--67, 2007.

\bibitem{zhu2002}
L.~Zhu and C.~Peskin.
\newblock Simulation of flapping flexible filament in a flowing soap film by
  the immersed boundary method.
\newblock {\em J. Comput. Phys.}, 179:452--468, 2002.

\bibitem{zhu2003}
L.~Zhu and C.~Peskin.
\newblock Interaction of two flapping filaments in a flowing soap film.
\newblock {\em Phys. Fluids}, 15:1954--1960, 2003.

\bibitem{bergou2007}
A.~Bergou, S.~Xu, and Z.~J. Wang.
\newblock Passive wing pitch reversal in insect flight.
\newblock {\em J. Fluid Mech.}, 591:321--337, 2007.

\bibitem{toomey2008}
J.~Toomey and J.~D. Eldredge.
\newblock Numerical and experimental study of the fluid dynamics of a flapping
  wing with low-order flexibility.
\newblock {\em Phys. Fluids}, 20:073603, 2008.

\bibitem{michelin2009a}
S.~Michelin and S.~G. Llewellyn~Smith.
\newblock An unsteady point vortex method for coupled fluid-solid problems.
\newblock {\em Theor. Comp. Fluid Dyn.}, 2009.
\newblock (in press).

\bibitem{michelin2008}
S.~Michelin, S.~G. Llewellyn~Smith, and B.~J. Glover.
\newblock Vortex shedding model of a flapping flag.
\newblock {\em J. Fluid Mech.}, 617:1--10, 2008.

\bibitem{alben2008c}
S.~Alben.
\newblock Optimal flexibility of a flapping appendage in an inviscid fluid.
\newblock {\em J. Fluid Mech.}, 614:355--380, 2008.

\bibitem{cortelezzi1993}
L.~Cortelezzi and A.~Leonard.
\newblock Point vortex model of the unsteady separated flow past a
  semi-infinite plate with transverse motion.
\newblock {\em Fluid Dyn. Res.}, 11:263--295, 1993.

\bibitem{cortelezzi1995}
L.~Cortelezzi.
\newblock On the unsteady separated flow past a semi-infinite plate. exact
  solution of the {B}rown and {M}ichael model, scaling and universality.
\newblock {\em Phys. Fluids}, 7:526--529, 1995.

\bibitem{brownmichael1954}
C.~E. Brown and W.~H. Michael.
\newblock Effect of leading edge separation on the lift of a delta wing.
\newblock {\em J.~Aero. Sci.}, 21:690--694 \& 706, 1954.

\bibitem{rott1956}
N.~Rott.
\newblock Diffraction of a weak shock with vortex generation.
\newblock {\em J.~Fluid Mech.}, 1:111--128, 1956.

\bibitem{jones2003}
M.~A. Jones.
\newblock The separated flow of an inviscid fluid around a moving plate.
\newblock {\em J.~Fluid Mech.}, 496:405--441, 2003.

\bibitem{alben2008}
S.~Alben and M.~J. Shelley.
\newblock Flapping states of a flag in an inviscid fluid: bistability and the
  transition to chaos.
\newblock {\em Phys. Rev. Lett.}, 100:074301, 2008.

\bibitem{saffman1992}
P.~G. Saffman.
\newblock {\em Vortex Dynamics}.
\newblock Cambridge University Press, 1992.

\bibitem{gorelov2008}
D.~N. Gorelov.
\newblock Calculation of pressure on an airfoil contour in an unsteady
  separated flow.
\newblock {\em J. Appl. Mech. Tech. Phys.}, 49:437--441, 2008.

\bibitem{alben2008d}
S.~Alben.
\newblock The flapping-flag instability as a non-linear eigenvalue problem.
\newblock {\em Phys. Fluids}, 20:104106, 2008.

\bibitem{lamb1932}
H.~Lamb.
\newblock {\em Hydrodynamics}.
\newblock Dover, New York, 6th edition, 1932.

\bibitem{kochin1964}
N.~E. Kochin, I.~A. Kibel, and N.~V. Roze.
\newblock {\em Theoretical Hydromechanics}.
\newblock Interscience Publishers, New York, 1964.

\bibitem{kornecki1976}
A.~Kornecki, E.~H. Dowell, and J.~O'Brien.
\newblock On the aeroelastic instability of two-dimensional panels in uniform
  incompressible flow.
\newblock {\em J. Sound Vib.}, 47:163--178, 1976.

\bibitem{eloy2007}
C.~Eloy, C.~Souilliez, and L.~Schouveiler.
\newblock Flutter of a rectangular plate.
\newblock {\em J. Fluids Struct.}, 23:904--919, 2007.

\bibitem{williamson1988}
C.~H.~K. Williamson and A.~Roshko.
\newblock Vortex formation in the wake of an oscillating cylinder.
\newblock {\em J. Fluids Struct.}, 2:355--381, 1988.

\bibitem{michelin2009b}
S.~Michelin and S.~G. Llewellyn~Smith.
\newblock Falling cards and flapping flags: understanding fluid-solid
  interactions using an unsteady point vortex model.
\newblock {\em Theor. Comp. Fluid Dyn.}, 2009.
\newblock (in press).

\bibitem{wang2007}
Z.~J. Wang and D.~Russell.
\newblock Effect of forewing and hindwing interactions on aerodynamic forces
  and power in hovering dragonfly flight.
\newblock {\em Phys. Rev. Lett.}, 99:148101, 2007.

\bibitem{zhang2000}
J.~Zhang, S.~Childress, A.~Libchaber, and M.~Shelley.
\newblock Flexible filaments in a flowing soap film as a model for
  one-dimensional flags in a two-dimensional wind.
\newblock {\em Nature}, 408:835--839, 2000.

\bibitem{jia2007}
L.-B. Jia, F.~Li, X.-Z. Yin, and X.-Y. Yin.
\newblock Coupling modes between two flapping filaments.
\newblock {\em J. Fluid Mech.}, 581:199--220, 2007.

\bibitem{jia2008}
L.-B. Jia and X.-Z. Yin.
\newblock Passive oscillations of two tandem flexible filaments in a flowing
  soap film.
\newblock {\em Phys. Rev. Lett.}, 100:228104, 2008.

\bibitem{ristroph2008}
L.~Ristroph and J.~Zhang.
\newblock Anomalous hydrodynamic drafting of interacting flapping flags.
\newblock {\em Phys. Rev. Lett.}, 101:194502, 2008.

\bibitem{bisplinghoff1955}
R.~L. Bisplinghoff, H.~Ashley, and R.~L. Halfman.
\newblock {\em Aeroelasticity}.
\newblock Addison--Wesley, 1955.

\bibitem{theodorsen1935}
T.~Theodorsen.
\newblock General theory of aerodynamic instability and the mechanism of
  flutter, 1935.
\newblock NACA Report 496.

\bibitem{abramowitz1964}
M.~Abramowitz and I.~A. Stegun.
\newblock {\em Handbook of Mathematical Functions with Formulas, Graphs, and
  Mathematical Tables}.
\newblock Dover, New York, 1964.

\bibitem{boyd2001}
J.~P. Boyd.
\newblock {\em Chebyshev and Fourier Spectral Methods}.
\newblock Dover, New York, 2nd edition, 2001.

\end{thebibliography}
\end{document}